\documentclass[11pt,pra]{revtex4}
\usepackage{amsmath}
\usepackage{amssymb}
\usepackage{bbm}
\usepackage{graphicx}

\newcommand{\tr}{\operatorname{Tr}}

\renewcommand{\vec}[1]{\boldsymbol{#1}}

\begin{document}
\title{Quantum Quenches in the Luttinger model
and its close relatives}
\author{M. A. Cazalilla}
\email{miguel.cazalilla@gmail.com}
\affiliation{Department of Physics, National Tsing Hua University, and National Center for Theoretical Sciences (NCTS), Hsinchu City, Taiwan}
\author{Ming-Chiang Chung}
\email{mingchiangha@phys.nchu.edu.tw}
\affiliation{Department of Physics, National Chung Hsing University, Taichung, Taiwan}
\begin{abstract}
A number of results on quantum quenches in the Luttinger and related models are surveyed with emphasis on post-quench correlations. For the Luttinger model and initial gaussian states, we discuss both sudden and smooth quenches of the interaction and the emergence of a steady state described by a generalized Gibbs ensemble. Comparisons between analytics and numerics, and the question of universality or lack thereof are also discussed. The relevance of the theoretical results to current and future experiments in the fields of ultracold atomic gases and mesoscopic systems of electrons is also briefly touched upon. Wherever possible, our approach is pedagogical and self-contained. This work is dedicated to the memory of our colleague Alejandro Muramatsu.  
\end{abstract}
\date{\today}
\maketitle

\tableofcontents

\section{Introduction}

 The study of non-equilibrium dynamics in isolated many-particle systems has become a very active research area in recent years~\cite{editorial,review_polkovnikov,review_rigol}. Many review articles have been devoted to various aspects of it~\cite{review_polkovnikov,review_rigol}.  This article focuses on a specific topic which is concerned with the quench dynamics of the Luttinger and related models. Even with this constraint in mind, the number of results
that have continued to appear since one of us~\cite{cazalilla2006} studied the quench of the interaction 
in this model in 2006 is fairly large. 

  Almost a decade has gone by, and with a certain perspective,
we have tried to provide a historical and personal account of how some of the ideas developed and what the concerns of the community at the time were. Of course, we have attempted to survey some of the most interesting developments in recent years, while providing a (hopefully) pedagogical introduction to the subject. This has forced us to make many choices in order to render
the article as self-contained and coherent as possible.  For this reason,  we would like to stress from the beginning that this work is far from perfect and cannot be considered comprehensive. When undertaking the task of surveying the field, we have tried our best to overcome our personal biases, however difficult this may be. But as humans facing space and time constraints, we may have ended up leaning towards what we know and understand best. Not surprisingly, this also overlaps strongly with our own work in the field.

  Nonetheless, we hope that this article will serve as a good starting point (and even as an inspiration!) for those students and non-experts willing learn about this fascinating subject. At the same time,  we apologize in advance to all the experts who, after going through the manuscript, find that we did not properly represent the most interesting aspects of their work, or those whose work has been (unintentionally) omitted. Hopefully some of those mistakes can be corrected in the future. Without further ado, let us get started.  The rest of the paper is divided in six sections. In the first two, we deal with the dynamics of the post-quench correlations in sudden and smooth quantum quenches. Section~\ref{sec:entgge} discusses the generalized Gibbs ensemble describing the asymptotic long-time state following a sudden quench, and its relation to the ideas of quantum entanglement. In section~\ref{sec:other}, we briefly survey some of the results obtained for other models that are related to the Luttinger model. Section~\ref{sec:exp} discusses the relevance of the results for experiments both with quantum gases and in mesoscopic physics.  Finally, in section \ref{sec:summ},  we provide our conclusions and an outlook.  From this point  on, we shall work in units where $\hbar = k_B = 1$.

\section{The Luttinger model in equilibrium: A brief history} \label{sec:history}

 Luttinger~\cite{luttinger1963} introduced the model that bears his name in 1963 as an example of an exactly solvable model of intreracting spinless 
fermions. However, the  solution that he obtained for his own the model  was not entirely correct, as it was shown shortly thereafter 
by Mattis and Lieb~\cite{mattislieb1965}.  

 Luttinger's assumptions included a linear dispersion for the fermions~\cite{luttinger1963}. However, since such a dispersion can take arbitrarily large negative values,  in order to obtain a physically sensible model with a spectrum bounded from below, Luttinger had to occupy all single-particle levels with negative kinetic energy with an infinite number of fermions. In other words,  the ground state of Luttinger's model is a `Dirac sea'. This was quite a departure from the non-relativistic models studied in the theory of quantum many-particle systems up to that point. In those models, such as the gas of interacting fermions with a parabolic dispersion,  the single-particle dispersion is bounded from below. Thus,   the ground state is a Fermi sea containing a finite number of fermions.  By contrast the number of particles in the Dirac sea is infinite and  Luttinger's model is indeed a quantum field theory in disguise. Indeed, in particle physics, the Lorentz-invariant version of Luttinger's model is known as the Thirring model.  
 
 The above observations were made by Mattis and Lieb~\cite{mattislieb1965}, who emphasized that the Dirac-sea character of the ground state has deep consequences for the structure of the Hilbert space and its operator content.  In particular, Luttinger had used a transformation to map the interacting model onto the non-interacting one~\cite{luttinger1963}.  The transformation appears to be  canonical but in reality is 
not~\cite{mattislieb1965}.  Indeed,  according to an earlier observation by Schwinger~\cite{schwinger1959}, the requirement of a Dirac sea makes the commutation relation of certain operators, such as the density, non-vanishing i.e. ``anomalous''. This happens independently of whether such operators commute in their first quantized form that applies to systems consisting of a finite number of particles.
 
 After discussing the structure of the (non-interacting) ground state, let us consider the form of the Hamiltonian. In the notation that we shall be following in the rest of the article, the second quantized  Hamiltonian can be written as the sum of three terms, i.e. $H_{LM} = H_0 + H_{2} + H_{4}$, where $H_0$ is the kinetic energy of the fermions with linear dispersion:
\begin{equation}
H_0 = \sum_{p} v_F p  \left[ :\psi^{\dag}_R(p) \psi_{R}(p): + :\psi^{\dag}_L(p) \psi_L(p):\right]\:.    
\end{equation}
In this expression $v_F$ is the Fermi velocity. The term $H_2+H_4$ describes the interactions:
\begin{align}
H_{2} &= \frac{2\pi }{L}\sum_{pkq} g_2(q)   :\rho_R(q) \rho_L(q):\, , \\
H_{4} &= \frac{\pi}{L}\sum_{pkq} g_4(q) : \left[ \rho_R(q) \rho_{R}(-q) + \rho_L(q)\rho_L(-q) \right]: \, .
\end{align}
In the above equations the operators $\psi_{\alpha}(p)$ ($\psi^{\dag}_{\alpha}(p)$)  annihilate  (create) fermions with momentum $p$ and chirality (i.e direction of motion) $\alpha = R, L$, and obey $\{ \psi_{\alpha}(p), \psi^{\dag}_{\alpha^{\prime}}(p^{\prime}) \} = \delta_{\alpha\alpha^{\prime}} \delta_{p,p^{\prime}}$, anti-commuting otherwise. For use further below, it is also useful to define the Fermi field operator:
\begin{equation}
\psi_{\alpha}(x) = \frac{1}{\sqrt{L}} \sum_{p} e^{i s_{\alpha} p x} \psi_{\alpha}(p),
\end{equation}
where $s_{R} = -s_{L} = 1$. In order to avoid a degenerate ground state, we assume the field operators to obey anti-periodic boundary conditions, i.e. $\psi_{\alpha}(x+L) = -\psi_{\alpha}(x)$, i.e. $p = \frac{2\pi}{L} \left( n + \frac{1}{2}\right)$, $n$ being an integer. The normal ordering of the operator $O$ is defined as $:O:\: = \: O -\langle 0 | O | 0\rangle$, where $|0\rangle$ is the ground state of the \emph{non-interacting} system.  In Luttinger's original model, the functions $g_{2}(q) = g_{4}(q)$, but in modern literature it has become
standard to treat them as different. It is also assumed that the interactions have a characteristic range, $R$, beyond which the decay to zero in real space. In terms of the Fourier components $g_2(q)$ and $g_4(q)$, this means that these functions rapidly vanish for $q \gg R^{-1}$. We shall also assume that they are free of singularities as $q\to 0$.

It was pointed out by Mattis and Lieb that the second-quantized density operators  $\rho_{\alpha}(q)  = \sum_{p}:\psi^{\dag}_{\alpha}(p+q)\psi_{\alpha}(p):$  satisfy the following algebra~\cite{mattislieb1965,bosonization_new,giamarchi_book}:
\begin{equation}
\left[ \rho_{\alpha}(q), \rho_{\alpha^{\prime}}(q^{\prime})\right] =  \frac{q L}{2\pi} \delta_{q+q^{\prime}} \delta_{\alpha\alpha^{\prime}}. \label{eq:km}
\end{equation}
In modern literature, this algebra  is known as the Abelian [U$(1)$] Kac-Moody (KM) algebra. It was  Mattis and Lieb's realization that the KM algebra is the key to the exact solubility of the model. This is because it is possible to rewrite the KM algebra in terms of the operators:
\begin{align}
a(q) &= -i \sqrt{\frac{2\pi}{|q|L}}\left[ \theta(q) \rho_{R}(-q)  -\theta(-q)  \rho_{L}(q)\right],\\
a^{\dag}(q) &=  i \sqrt{\frac{2\pi}{|q|L}}\left[ \theta(q) \rho_{R}(q)  -\theta(-q)  \rho_{L}(-q)\right],
\end{align}
such that $\left[a(q), a^{\dag}(q^{\prime}) \right] = \delta_{q,q^{\prime}}$ and commute otherwise, as corresponds to canonical bosons. In addition, Mattis and Lieb rediscovered an \emph{exact} result obtained by Jordan~\cite{jordan_neutrino} in the context of his neutrino theory of light, which states that the kinetic energy of the fermions with linear dispersion can be written as:
\begin{equation}
H_0 = \sum_{q\neq 0}  v_F |q| \: a^{\dag}(q) a(q).\label{eq:hamkins}
\end{equation}
Thus, since the interactions $H_2$ and $H_4$ are quadratic in the density operators, which means they are also quadratic in $a(q)$ and $a^{\dag}(q)$, we  obtain with a quadratic Hamiltonian in the bosonic basis, which  can be diagonalized by means of a canonical (Bogoliubov) transformation:
\begin{equation}
a(q) = \cosh \varphi(q) b(q) + \sinh \varphi(q) b^{\dag}(-q),\label{eq:bogoleq}
\end{equation}
where  $\left[a(q), a^{\dag}(q^{\prime}) \right]  =\delta_{q,q^{\prime}}$.
The Bogoliubov angle $\varphi(q)$ is determined from the equation:
\begin{equation}
\mathrm{tanh} \left( 2\varphi(q)\right) = \frac{g_2(q)}{v_F + g_4(q)}.
\label{eq:bogolangle}
\end{equation}
Therefore, the Hamiltonian of the interacting system, $H$, is diagonal in terms of the new bosonic operator basis $\{b(q), b^{\dag}(q)\}$:
\begin{equation}
H_{LM} = \sum_{q\neq 0}  v(q) |q| \: b^{\dag}(q) b(q).~\label{eq:hambos}
\end{equation}
where the boson velocity is given by:
\begin{equation}
v(q) = \sqrt{(v_F + g_4(q))^2 -(g_2(q))^2}.\label{eq:veloc}
\end{equation}
Mattis and Lieb's solution of the Luttinger model (LM) provided the first concrete 
example of an interacting Fermi system exhibiting an excitation spectrum that strongly deviates from Landau's normal ``Fermi-liquid'' paradigm. The spectrum of the model, as shown in Eq.~\eqref{eq:hambos}, consists of collective, plasmon-like, bosonic modes known as Tomonaga bosons. These\ \emph{bosonic} elementary excitations are quite unlike the fermionic quasi-particles that describe the low-lying states of Fermi liquids. 

 Perhaps the most striking signature of the failure of the LM to conform to the framework of normal Fermi liquids can be observed in the momentum distribution. Mattis and Lieb noticed that, in the thermodynamic limit (i.e. for $L \to +\infty$) instead of the characteristic discontinuity at the Fermi momentum $p_F$,
the momentum distribution of the LM exhibits a much weaker, power-law singularity:
\begin{equation}
n(p) = \int dx \, e^{-ip x} C^{GS}_{\psi_R}(x) \approx \frac{1}{2} + |p-p_F|^{\gamma^2_{\mathrm{eq}}} \: \mathrm{sgn}(p-p_F),  \label{eq:nk}
\end{equation}
for $p\approx p_F$; the exponent  $\gamma^2_{\mathrm{eq}} =\cosh\left( 2 {\varphi(q=0)}  \right) - 1$ depends on the details of the interaction;  $C^{GS}_{\psi_R}(x) = \langle \psi^{\dag}_{R}(x)\psi_R(0) \rangle$ is the single-particle density matrix (the expectation value $\langle \ldots \rangle$ is taken over the ground state of the \emph{interacting} system).   Mattis and Lieb were able to obtain this result by a method equivalent to bosonization~\cite{bosonization_old,bosonization_new}. Here, we shall recall the main identities and results of this method, referring the interested reader to the vast available literature on the subject~\cite{bosonization_new,giamarchi_book,cazalilla2004}. The method relies on the following identity:
\begin{equation}
\psi_{\alpha}(x) = \frac{\eta_{\alpha}}{\sqrt{2\pi a}} e^{i s_{\alpha} \phi_{\alpha}(x)}\label{eq:bosonization}.
\end{equation}
This allows to express the Fermi fields in terms of the boson field ($s_{R} = -s_L = 1$):
\begin{align}
\phi_{\alpha}(x) &= s_{\alpha} \phi_{0\alpha} + \frac{2\pi x}{L} N_{\alpha} + \Phi_{\alpha}(x) + \Phi^{\dag}_{\alpha}(x), \label{eq:phib}\\
\Phi_{\alpha}(x) &= \sum_{q >  0}  \left(\frac{2\pi}{q L} \right)^{1/2} e^{-a_0 q/2}  e^{i q x} \, a(s_{\alpha} q), \notag
\end{align}
where $a_0$ is a short distance cut-off and $\{\eta_{\alpha},\eta_{\alpha^{\prime}}\} = \delta_{\alpha\alpha^{\prime}}$, which ensures the anti-commutation between fermions of different chirality.  The operators $\phi_{0\alpha}$ and $N_{\alpha} = \sum_{p} :\psi^{\dag}_{\alpha}(p) \psi_{\alpha}(p):$ are a canonically conjugate pair (i.e. $\left[ N_{\alpha},\phi_{0\alpha^{\prime}}\right] = i \delta_{\alpha,\alpha^{\prime}}$). Thus, 
\begin{equation}
C^{\mathrm{GS}}_{\psi_R}(x) = \frac{1}{2\pi a} \langle e^{i\phi_{R}(x)} e^{-i \phi_R(0)}\rangle  = 
Z^{\mathrm{GS}}(x) C^{(0)}_{\psi_R}(x),
\end{equation}
where
\begin{align}
C^{(0)}_{\psi_R}(x) &=  \frac{e^{ip_F x}}{2i L \sin\left[\pi (x+ia_0)/L\right]},\\ 
Z^{\mathrm{GS}}(x) &=  \left(\frac{R}{d(x|L)}\right)^{\gamma^2_{\mathrm{eq}}}
\end{align}
In the above expressions, $C^{(0)}_{\psi_R}(x)$  is the \emph{non-interacting} single-particle density matrix and $d(x|L) = L |\sin(\pi x/L)|/\pi$ the \emph{cord} function and $\gamma^2_{\mathrm{eq}}$ is the equilibrium exponent that has been introduced under Eq.~\eqref{eq:nk} above. Another
correlation function of interest is the density correlation function. In terms of the 
boson field, the density operator $\rho_{\alpha}(x) =  \partial_{x}\phi_{\alpha}(x)/2\pi$~\cite{bosonization_new,giamarchi_book}, and therefore
\begin{equation}
C^{\mathrm{GS}}_{\rho_R}(x) = \langle \rho_R(x) \rho_R(0)\rangle = -\frac{e^{-2\varphi(0)}}{4\pi^2} \left[\frac{1}{d(x|L)} \right]^2,
\end{equation}
which also exhibits an algebraic decay with distance, but with a exponent that is independent of the interaction (although the pre-factor is not). 

 As pointed out by Mattis and Lieb~\cite{mattislieb1965}, the solution of the LM shares many interesting properties with the approximate solution of the one-dimensional electron gas obtained by Tomonaga~\cite{tomonaga1950} in 1950. The striking resemblance was to become more and more important in the course of time. Indeed, after Luttinger's and  Mattis and Lieb's seminal contributions,  the exotic properties of the model turned out to be more than a just a mathematical curiosity. Beginning in the 1970s, the study of one-dimensional interacting systems started to attract an increasingly large amount of attention motivated by the advances in materials synthesis and spurred by Little's proposal~\cite{little1964} for a new class of organic high-temperature superconductors based on highly anisotropic materials made up of quasi-one dimensional metallic molecules. Ground breaking work along in this direction was done by Luther and Emery~\cite{lutheremery1974} by extending the Luttinger model to spinful fermions and finding that, when backscattering processes are taken into account,  the system develops spectral gap to spin excitations but remains gapless for charge excitations. Such system, currently known as the Luther-Emery liquid, is the canonical example of a one-dimensional superconductor. In addition, Luther in collaboration with Peschel~\cite{lutherpeschel1975} provided the first crucial insights into the fundamental observation that the low-temperature behavior of the LM is universal and applies to a general class of 1D models. Building upon the earlier work by Luther and Peschel on the anisotropic Heisenberg XYZ spin-chain model, and in the spirit of Landau's Ferm liquid theory, Haldane~\cite{haldane_tll,haldanejpc} coined the name ``Luttinger liquids'' for this new universality class, which encompasses a large class of one-dimensional models of fermions~\cite{mattislieb1965}, bosons~\cite{haldane1981}, and spins~\cite{lutherpeschel1975,haldane_tll}.
 
  Indeed, in the language of the renormalization group~\cite{rg_shankar,rg_book},  the LM turns out to be a fixed point Hamiltonian for this universality class. The fixed point Hamiltonian of a TLL can be written in terms of the (total) density $\phi = (\phi_R + \phi_L)/2$ and phase $\theta = (\phi_R - \phi_L)/2$, as follows:
\begin{equation}
H_{TLL} = \frac{v}{2\pi} \int dx \left[ K^{-1} \left(\partial_x \phi \right)^2 + K \left(\partial_x \theta \right)^2 \right],  \label{eq:tllham}
\end{equation}
where the Tomonaga boson velocity is $v  = v(q = 0)$ (cf. Eq.~\ref{eq:veloc}), the so-called Luttinger parameter is $K = e^{-2\varphi(q = 0)}$ in terms of Bogoliubov rotation angle at $q = 0$ (cf. Eq.~\ref{eq:bogolangle}).  Note that the  density stiffness is proportional to $v K^{-1}$ and the phase stiffness is related to $v K$. In Galilean-invariant systems $v K =  \rho_0/m$, where $\rho_0$ is the 
particle density~\cite{haldane1981,giamarchi_book,cazalilla_rmp}.   The Hamiltonian in Eq.~\eqref{eq:tllham} can be diagonalized and brought to a form similar to Eq.~\eqref{eq:hambos}, with
$v(q)$ replaced by $v = v(q = 0)$. Thus,  the low-energy excitation spectrum 
is completely exhausted  by the Tomonaga bosons. In recognition to Tomonaga's pioneering contributions, the universality class  has been renamed as ``Tomonaga-Luttinger liquids' (TLLs).     The defining properties of the systems in the TLL universality class are a linear dispersion of the Tomonaga bosons, and the power-law correlations at zero temperature. 
The exponents of the power-laws  are parametrized by $K$. Notice that the previous observations
apply to systems for which the interactions at $q = 0$ 
are not singular, which implies that both $v$ and $K$ are
finite as $q \to 0$. This is not the case when the interactions are long range such like the case of the
Coulomb interation. We refer the interested reader to
e.g. Ref.~\cite{giamarchi_book} (and references therein) 
for an  account of how the correlations are modified in this case. 

\section{Dynamics after a sudden quench}\label{sec:dyn}

\subsection{Introduction and historical context}\label{subsec:prel}

In 2006, one of us (MAC) attended a workshop at the Max Planck Institute for Complex Systems in Dresden. The workshop, co-organized by Alejandro Mumamatsu, was devoted to non-equilibrium dynamics in interacting Systems. At the workshop, Rigol reported on the results of his ground-breaking work with Dunjko, Yurovskii, and Olshanii,  motivated by experiments performed in Weiss' group~\cite{kinoshita2006}. In their work,  Rigol and coworkers showed that gas of lattice hardcore bosons in 1D (also known as the XX model) does not thermalize to the standard Gibbs ensemble when prepared in an initial state that is not an eigenstate of the Hamiltonian. In their study, they provided convincing numerical evidence that the system instead relaxes to a `generalized' Gibbs ensemble, which is constructed by using a particular set of integrals of motion of the system (see section~\ref{sec:entgge} for a description).  

 Following Rigol's report at the workshop, ensuing discussions with Muramatsu and Rigol provided enough motivation for one of us (MAC) to search for an analytically-solvable example of such peculiar behavior.  Out of  the several possible candidates, the Luttinger model seemed the most natural choice. However, several obstacles had to be surmounted. In Rigol \emph{et al.}'s work, two kinds of initial states had been considered: In one of them, it was assumed the hardcore boson gas  is initially trapped in a smaller box and suddenly released into a larger box. The second choice was  motivated  by one of the experiments reported in Ref.~\cite{kinoshita2006},  and considered that a bi-periodic potential is applied in the initial state of the lattice boson gas and suddenly switched off at start of the evolution~\cite{rigoletal2006}. 
 
 Neither of the above choices of initial states considered seemed analytically tractable in the case of the LM because they break translational invariance (however, see remarks at the end of section~\ref{subsec:LMEP}). Instead, what seemed most natural and amenable to analytical calculation was to assume that the interaction between the fermions, described by the terms $H_2+H_4$ in the Hamiltonian, is suddenly switched-on at the start of the time evolution. 
An attractive feature of such a quench is that the eigenstates of the non-interacting system can be still described in terms of fermionic eigenmodes. On the other hand,  the eigenstates of $H_{LM} = H_0 + H_2 +H_4$ are described in terms of Tomonaga bosons. Thus,  the quench of the interaction  allows to study how the fermionic  features of the spectrum are dynamically destroyed. 
 
 Assuming that the system is prepared in the non-interacting ground state, $| 0 \rangle$ and it evolves according to $H_{LM} = H_0 + H_2 + H_4$ for all times $t > 0$ is equivalent to a \emph{quantum quench} of the interaction. The term ``quantum quench'' was introduced by Cardy and Calabrese~\cite{cardycalabrese2006} in their 2006 pioneering work (see also this volume) where they studied the evolution of correlations when the system is suddenly driven from a off-critical to a critical state. The main difference with situation considered by Cardy and Calabrese is that  the initial state in 
Ref.~\cite{cazalilla2006} lacks any characteristic length scale, i.e. it is a critical state.

\subsection{Dynamics of the Luttinger model following a quench of the interaction}\label{sec:corrdyn}

 The solution of the quench of the interaction in the LM can be obtained with minimal use of formalism in the operator language\footnote{Keldysh aficionados are referred to e.g. Refs.~\cite{mitra,Mirlin}, where the same solution is obtained using the non-equilibrium Green's function formalism.}. The solution takes the form of a  canonical transformation relating the bosonic operators at times $t  = 0$ and $t  > 0$:
\begin{equation}
a(q,t) = e^{i H t} a(q) e^{-iH t} = f(q,t) a(q) + g^*(q,t) a^{\dag}(-q), \label{eq:canont}
\end{equation}
where the time-dependent (Bogoliubov) coefficients read:
\begin{align}
f(q,t) &= \cos \left( v(q) |q| t \right)  - i \cosh\left( 2\varphi(q)\right)\:  \sin \left(v(q) |q| t\right)  \\
g(q,t) &= i \sinh \left( 2\varphi(q) \right) \: \sin \left( v(q) |q| t \right).
\end{align}
Note that $f(q,t=0) = 1$ and $g(q,t=0) = 0$, in agreement with the initial state of the system being the non-interacting ground state, $|0\rangle$, which is annihilated by $a(q)$ for all $q\neq 0$. In addition, 
Eq.~\eqref{eq:canont} shows that the time evolution with the interacting Hamiltonian $H_{LM}$ alters, in a time-dependent way, the entanglement of 
the modes at opposite momenta $q$ and $-q$, which is a consequence of the translational invariance of both the Hamiltonian and the initial state. We shall return to this point in Section~\ref{sec:entgge}.
\begin{figure}
\centering
\includegraphics[scale=0.47]{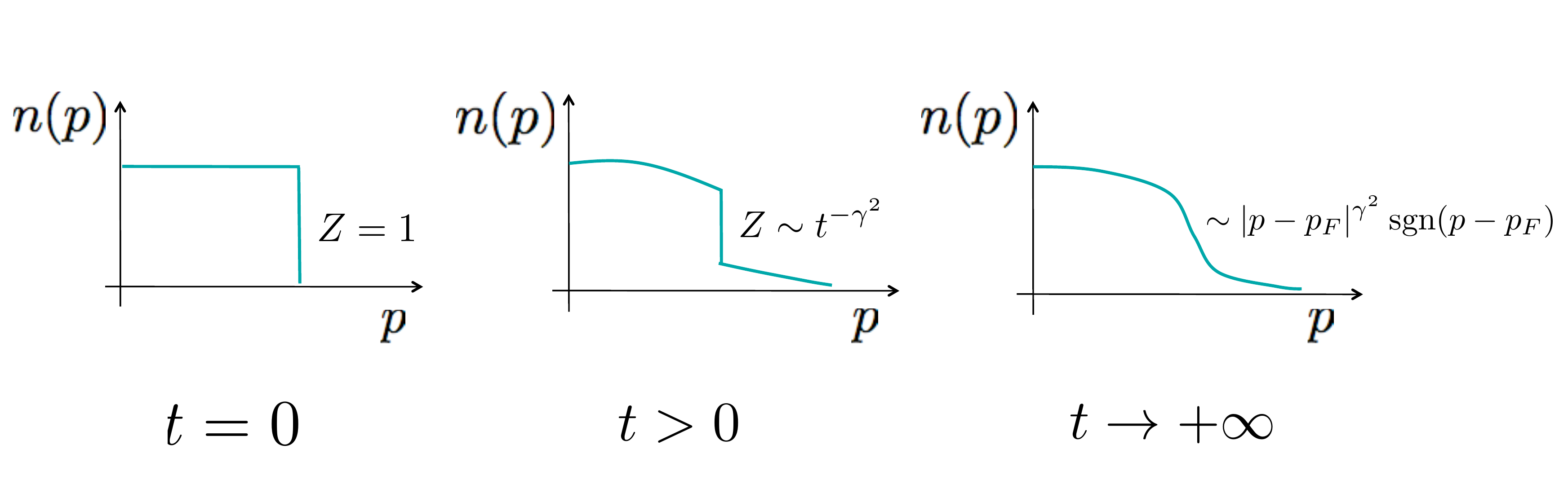}
\caption{Schematic evolution of the momentum $n(p)$ distribution of the Luttinger model  in the neighborhood of the Fermi momentum, $p_F$, following
a quench of the interaction. The non-equilibrium exponent $\gamma^2 = (K^2 + K^{-2}-2)/4$, where $K$ is the Luttinger parameter, which parametrizes the correlations in the ground state, is different from the equilibrium exponent, $\gamma^2_{\mathrm{eq}} = (K + K^{-1}-2)/2$ }
\label{fig:fpt}
\end{figure}
In order to obtain the time evolution of correlations from the initial (non-interacting) state and study how the fermionic features of the non-interacting
LM are wiped out, let us consider the instantaneous single-particle density matrix~\cite{cazalilla2006}:
\begin{equation}
C_{\psi_R}(x,t) = \langle 0 | e^{iH_{LM}t} \psi^{\dag}_R(x)\psi(0) e^{-i H_{LM}t} | 0\rangle =  \langle 0 |  \psi^{\dag}_R(x,t)\psi_R(0,t)  | 0\rangle,\label{eq:opd}
\end{equation}
which can be computed using the bosonization formula, Eq.~\eqref{eq:bosonization}, together with Eq.~\eqref{eq:canont}. The calculation proceeds pretty much along the lines of the calculation in the equilibrium case~\cite{cazalilla2004,cazalilla_rmp,giamarchi_book,bosonization_new}, and  yields the following result in the scaling limit (i.e. for $d(x|L), d(2 vt|L) \gg R$):
\begin{align}
C_{\psi_R}(x,t) &=  Z(x,t) C^{(0)}_{\psi_R}(x),\label{eq:cpsi} \\
Z(x,t) &= \left[\frac{R}{d(x|L)} \right]^{\gamma^2} \left[\frac{d(x-2vt|L) d(x+2vt|L)}{d(2vt)} \right]^{\gamma^2/2}, 
\end{align}
Likewise, we can obtain the density correlations:
\begin{align}
C_{\rho_R}(x,t) &= \langle 0| e^{iH_{LM}t} \rho_R(x) \rho_R(0) 
e^{-iH_{LM}t} | 0\rangle = \langle 0|  \rho_R(x,t) \rho_R(0,t) | 0\rangle\\
 &= -\frac{1}{4\pi} \left\{ \frac{1+\gamma^2}{\left[ d(x|L)\right]^2} 
 - \frac{\gamma^2}{2\left[d(x-2vt|L) \right]^2} -  \frac{\gamma^2}{2\left[d(x+2vt|L) \right]^2}\right\}. \label{eq:rhor}
\end{align}
In the avobe expressions $v = v(q = 0)$ and $\gamma^2 = \sinh^2 \left(  2\varphi(q = 0) \right) = (K^{2}+K^{-2}-2)/4$, where $K = e^{2\varphi(q=0)}$ is the Luttinger parameter.  Note that since the correlation functions are periodic functions of time and will exhibit periodic recurrences with a period $t_L = L/2v$.  This is the consequence of the finite size of the system.  In the thermodynamic limit, $L\to +\infty$,  and therefore $t_L \to +\infty$. Thus, the cord functions yield power-laws by the replacement $d(z|L) \to |z|$.

 For $L\to +\infty$ (or for $|x|, 2vt \ll L$), the behavior of the above correlation functions exhibits two clearly distinct regimes. At short times, i.e. for $t \ll |x|/2v$, 
\begin{align}
C_{\psi_R}(x,t) &\simeq Z(t) C^{(0)}_{\psi_R}(x), \quad Z(t) \sim t^{-\gamma^2}\\
C_{\rho_R}(x,t) &\simeq -\frac{1}{4\pi^2 x^2} = C^{(0)}_{\rho_R}(x). 
\end{align}
That is, the correlations take a similar form to those of the non-interacting system. In the case of the single-particle density matrix, it decreases by an overall factor $Z(t)\sim t^{-\gamma^2}$. On the other hand,  in the long time regime, i.e. for $t \gg |x|/2v$, the correlations crossover to
\begin{align}
C_{\psi_R}(x,t) &\simeq  C^{(0)}_{\psi_R}(x) \left|\frac{R}{x} \right|^{\gamma^2},\\
C_{\rho_R}(x,t) &\simeq -\frac{1+\gamma^2}{4\pi^2 x^2}. \label{eq:denscor}
\end{align}
In this case, the correlations become qualitatively different from the initial state correlations. In particular, to leading order, they do not depend on time, which indicates that for $t \to  +\infty$ the system reaches a steady state.
We shall investigate this behavior more in detail further below, in section~\ref{sec:entgge}. However, at this point, it is worth pointing out that the existence of the two distinct correlation regimes separated by a time scale $t_x = |x|/2v$ has to do with the finite propagation velocity of the elementary excitations of the LM. Since the LM is a relativistic model in which  
the role of the speed of light is played by the boson velocity $v$, this can be expected. The time scale $t_x$ is thus related to the time it takes for the excitations in the initial state localized at two points a distance $|x|$ apart to overlap. This phenomenon has been termed 'light-cone effect'  by Cardy and Calabrese~\cite{cardycalabrese2006} in their study of quantum quenches starting from an off-critical state and ending in a critical state. In a more general framework, it is the consequence of the Lieb-Robison-type bounds~\cite{liebrobinson} for propagation of signals in systems with finite-range interactions.

The correlations in the steady state are different from those of the initial state and from those in the ground state of the LM. Despite the fact that the initial state is a complicated superposition of eigenstates of $H_{LM}$, the correlations are not thermal. If they were, they would exhibit an exponential decay beyond a characteristic length scale determined by the final (effective) temperature of the system $T_{f}$ (see e.g.~\cite{cazalilla2004,cazalilla_rmp,giamarchi_book,bosonization_new} and references). To see this, let us  consider an ``infinitesimal'' quench in which an very weak interaction (i.e. $\varphi(q = 0) \ll 1$) is switched on. Thus, the energy of the initial state differs  from the energy of the LM  ground state (taken to be zero) by an infinitesimal amount,
\begin{align}
\frac{\Delta E}{L} = \frac{1}{L}\langle 0 | H_{LM} | 0\rangle &= \frac{1}{L}\sum_{q\neq 0} v(q) |q|  \langle 0| b^{\dag}(q) b(q) | 0\rangle 
%\\&= \frac{1}{L}\sum_{q\neq 0}  v(q) |q| \sinh^2 \varphi(q)
\approx  \frac{v}{\pi R^2} \sinh^2 \varphi(q =0) \simeq  \frac{v}{\pi R^2} \varphi^2(q=0),
\end{align}
which means that only low-lying excited states should be involved in the evolution following the quench. Thus, we can use the low-temperature form of the free energy in the canonical Gibbs ensemble for the LM
(see e.g. Ref.\cite{mattislieb1965}) $F(T)/L \simeq  \pi^2 T^2_fv/6$. Hence, 
the corresponding final temperature, $T_{f}\propto \varphi(q=0)$. However, this is still finite, which means that, if the system relaxes to the canonical Gibbs ensemble,  the correlations should decay exponentially at long distances~\cite{bosonization_new,cazalilla2004}, e.g.:
\begin{equation}
C_{\psi_R}(x,t\to+\infty) \sim e^{-|x|/\xi(T_f)}, \label{eq:corrthermal}
\end{equation}
for $|x| \gg \xi(T_f)$  where $\xi(T_f) \propto v/T_f$. By contrast, the asymptotic  $t\to +\infty$ state following the quantum quench exhibits power-law correlations, as we have shown above. In other words, the system does not thermalize to the canonical Gibbsian ensemble. We will see in Sec.~\ref{sec:entgge} that this is because the LM is exactly solvable and relaxes instead to a generalized Gibbs ensemble. 

Before discussing the properties of the asymptotic state, let us discuss the consequences of the above results for the evolution of the momentum distribution.
As discussed in Sec.~\ref{sec:history}, the latter is the Fourier transform of the equilibrium single-particle density matrix. By analogy, we can define the instantaneous momentum distribution, 
\begin{equation}
n(p,t)  = \int dx \: e^{-ip x} C_{\psi_R}(x,t) = \left\{
 \begin{array}{cc}
Z(t)\: \theta(p_F-p) & \text{for finite}\: t\quad  (Z(t) \sim t^{-\gamma^2}),\\
\frac{1}{2}+ \mathrm{sgn}(p-p_F) |p_F-p|^{\gamma^2-1}  & \mbox{for } t\to +\infty.
 \end{array} \right. 
\end{equation}
The results on the left-hand side hold for $p\approx p_F$.  Fig.~\ref{fig:fpt} shows
the evolution of the momentum distribution schematically. The discontinuity
at the Fermi momentum $p_F$ decreases algebraically and  closes for $t \to +\infty$ becoming a weaker, power-law
singularity, with an exponent that differs from the exponent characterizing the
discontinuity at $p_F$ in the ground state.

Additional post-quench correlation functions were obtained by Iucci and Cazalilla in Ref.~\cite{iucci2009}. We just quote here their results for the correlations of the so-called vertex operators starting from the non-interacting ground state. For the ratio of the non-equilibrium to the initial state correlations, the following results hold in the thermodynamic limit: 
\begin{align}
\frac{C_{V^{2m}_{\phi}}(x,t)}{C^{(0)}_{V^{2m}_{\phi}}(x)} &= \frac{\langle 0|e^{2im\phi(x,t)} e^{-2im \phi(0,t)} |0\rangle}{\langle 0| e^{2im \phi(x)} e^{-2im\phi(0)}|0 \rangle} = \mathcal{A}^{\phi}_m  \left|\left(\frac{R}{2 v t}\right) \left(\frac{x^2- (2v t)^2}{x^2} \right) \right|^{m^2(K^2-1)/2}, \label{eq:cphi}\\
\frac{C_{V^m_{\theta}}(x,t)}{C^{(0)}_{V^m_{\theta}}(x)} &=  \frac{\langle 0|e^{im\theta(x,t)} e^{-im \theta(0,t)} |0\rangle}{\langle 0| e^{im \theta(x)} e^{-im\theta(0)}|0 \rangle} = \mathcal{A}^{\theta}_m \left|\left(\frac{R}{2 v t}\right) \left(\frac{x^2- (2v t)^2}{x^2} \right) \right|^{m^2(K^{-2}-1)/8},\label{eq:ctheta}
\end{align}
where $m$ is an integer. In the above expressions~\cite{cazalilla_rmp,cazalilla2004,bosonization_new,giamarchi_book},
\begin{align}
C^{(0)}_{V^{2m}_{\phi}}(x) &= \langle 0| e^{2im \phi(x)} e^{-2 im\phi(0)}|0 \rangle = \left| \frac{R}{x} \right|^{2 m^2},\\ 
C^{(0)}_{V^{m}_{\theta}}(x) &= \langle 0| e^{im \theta(x)} e^{-im\theta(0)}|0 \rangle = \left| \frac{R}{x} \right|^{m^2/4}.
\end{align}
Notice that  the usual equilibrium duality relations requiring  that if $\phi \to \theta$ then $K\to K^{-1}$ also hold for the post-quench correlations. Besides the above results, Iucci and Cazalilla also studied correlations at finite temperature, obtaining the momentum distribution  in  the asymptotic steady state at $t\to +\infty$. In Ref.~\cite{iucci2009}, a quantum quench in which the interactions are suddenly switched off was also studied. We refer the interested reader to section II.-C of Ref.~\cite{iucci2009} for the detailed form of the post-quench correlations in this  case. 

 In the derivation of the previous results,  it has been assumed that the interactions are long ranged but  not singular. 
This translates into $g_2(q)$  and $g_4(q)$  being regular 
functions of $q$ as $q \to 0$, which is necessary to ensure that $K$ and $v$ are both finite. This is not the case for the Coulomb interaction for which $g_2(q) = g_4(q) = V(q)/2\pi$. Here  $V(q) = 2e^2  K_0(q d)$ is the Fourier transform of the Coulomb potential,  $K_0(x)$ being the modified Bessel function and  $d$ a length scale of the order of the transverse dimensions system (recall that $K_0(x \ll 1) = \log\left(\frac{2}{x e^{\gamma}}\right) + \cdots$, where $\gamma = 0.5772156649\ldots$ is Euler's constant). The post-quench correlations for the LM with Coulomb interactions have been obtained by Nessi and Iucci~\cite{nessi2013}.  In what follows, we reproduce here their results for the single-particle density matrix, specializing to the case of a quench from the non-interacting system (which corresponds to setting $K_i = 1$ in their expressions). For the single-particle density matrix $C_{\psi_R}(x,t)$, the following expression for the factor $Z(x,t)$ in Eq.~\eqref{eq:cpsi} was obtained~\cite{nessi2013}:
\begin{align}
Z(x,t) &= \exp\left\{ -  \int^{+\infty}_0 \frac{dq}{q} \:  \sinh^2\left[ \varphi(q) \right] \left[1
-\cos\left(2 v(q) q t \right)\right] \left(1 - \cos q x\right)
\right\},\label{eq:corrcoul}
\end{align}
where 
$v(q) = \sqrt{v_F(v_F + 2 V(q)/\pi)}$  is the Tomonaga boson velocity, $v(q\ll 1/d) \sim v_F \log^{1/2}(1/qd)$; the  Bogoliugov rotation angle of the LM in the presence of Coulomb interactions follows from $\tanh 2 \varphi(q) = V(q)/(2\pi v_F + V(q))$. Asymptotic
expressions can be obtained from Eq.~\eqref{eq:corrcoul}. For instance,  the asymptotic long time 
limit reads~\cite{nessi2013}:
\begin{equation}
C_{\psi_R}(x,t \to +\infty) = C^{(0)}_{\psi_R}(x) e^{-\frac{g}{4}\log^2(x/d)},
\end{equation}
where $g = e^2/\pi v_F$, $e$ being the fundamental fermion charge. 
This form again differs from the equilibrium expression~\cite{giamarchi_book,nessi2013}. The intermediate time dynamics, however, is complicated by the divergence of the  Tomonaga boson velocity $v(q)$ as $q \to 0$, which leads to a non-linear light-cone effect~\cite{nessi2013}. Thus, for times fulfilling the condition: 
\begin{equation}
\frac{d}{v_F} \ll t \ll t_x  = \frac{x}{2v_F \sqrt{1+ 2g\log(d/x)}},
\end{equation}
the single-particle density matrix takes the form:
\begin{equation}
C_{\psi_R}(x,t) = C^{(0)}_{\psi_R}(x) e^{-\frac{g}{4}\log^2(2 v_F t/d)}
\end{equation}
with exponential accuracy. Hence, it follows that the discontinuity at $p_F$ in momentum distribution  decreases as $Z(t) \sim e^{-\frac{g}{4}\log^2(2 v_F t/d)}$
instead of the power-law $\sim t^{-\gamma^2}$ 
found for the LM with non-singular interactions.  
Asymptotic forms for other post-quench correlations, such as those of vertex operators,  were also obtained in Ref.~\cite{nessi2013}, and we refer the interested reader to the original article for the details. 
\subsection{Quest for universality: Quenches in models of the TLL class}\label{sec:quest}
In the previous subsection, we have reviewed the results obtained for
a sudden quench of the interaction in the LM. Since sudden quenches can potentially
drive the system far from equilibrium, there is no reason to expect that the results discussed above can be universal. Indeed, universality applies to the low-temperature, long distance and time correlations of systems in \emph{equilibrium} and it is borne out on the  ideas of the 
renormalization group~\cite{rg_shankar,rg_book}. According to the latter, the low-temperature properties of a system are rather insensitive to the microscopic 
details as it is the structure of the low-lying excited states. 
Therefore,  a sudden quantum quench that involves highly excited states
is not likely to yield correlations that are universal. 
 
 Nevertheless, since LM is a renormalization-group fixed point for the Tomonaga-Luttinger liquid universality class, there is much  interest in 
investigating to which extent correlations following a sudden quench
exhibit  universality. Analytical progress in this regard is particularly difficult. Therefore, in order to ascertain whether the correlations are independent or not of the microscopic details of the model, 
a number of  numerical and semi-numerical techniques have been deployed.

 In particular, the LM prediction for the dynamics of the discontinuity at $p_F$ in the momentum
distribution reviewed in the pervious section, has been numerically tested
by Karrasch, Rentrop, Schuricht, and Meden (KRSM)~\cite{karraschmedem2012} using  time-dependent DMRG~\cite{tddmrg}. KRSM considered a (sudden) quench of the interaction in the following lattice model:
\begin{equation}
H =  \sum_{m=1}^L \left[-\frac{1}{2}  \left(  c^{\dag}_{m+1}c_{m} + c^{\dag}_{m} c_{m+1} \right) + \sum_{m} \left( \Delta  n_m n_{m+1} + \Delta_2 n_{m} n_{m+2} \right)  \right],
\label{eq:modeldd2} 
\end{equation}
from the non-interacting system, i.e. starting
from the ground state of the XX model (i.e. $\Delta = \Delta_2 = 0$) to the interacting model with either $\Delta \neq 0$ or $\Delta\neq 0$ and $\Delta_2 \neq 0$. In the former case, the model is known as the XXZ model and it can be solved exactly using the Bethe-ansatz method (see e.g.~\cite{giamarchi_book,cazalilla_rmp} and references therein). However, when the term proportional to $\Delta_2$ is present, the model is no longer integrable in Bethe-ansatz sense. Yet, the results in both cases showed good agreement with the LM predictions.  
\begin{figure}
\centering
\includegraphics[scale=.8]{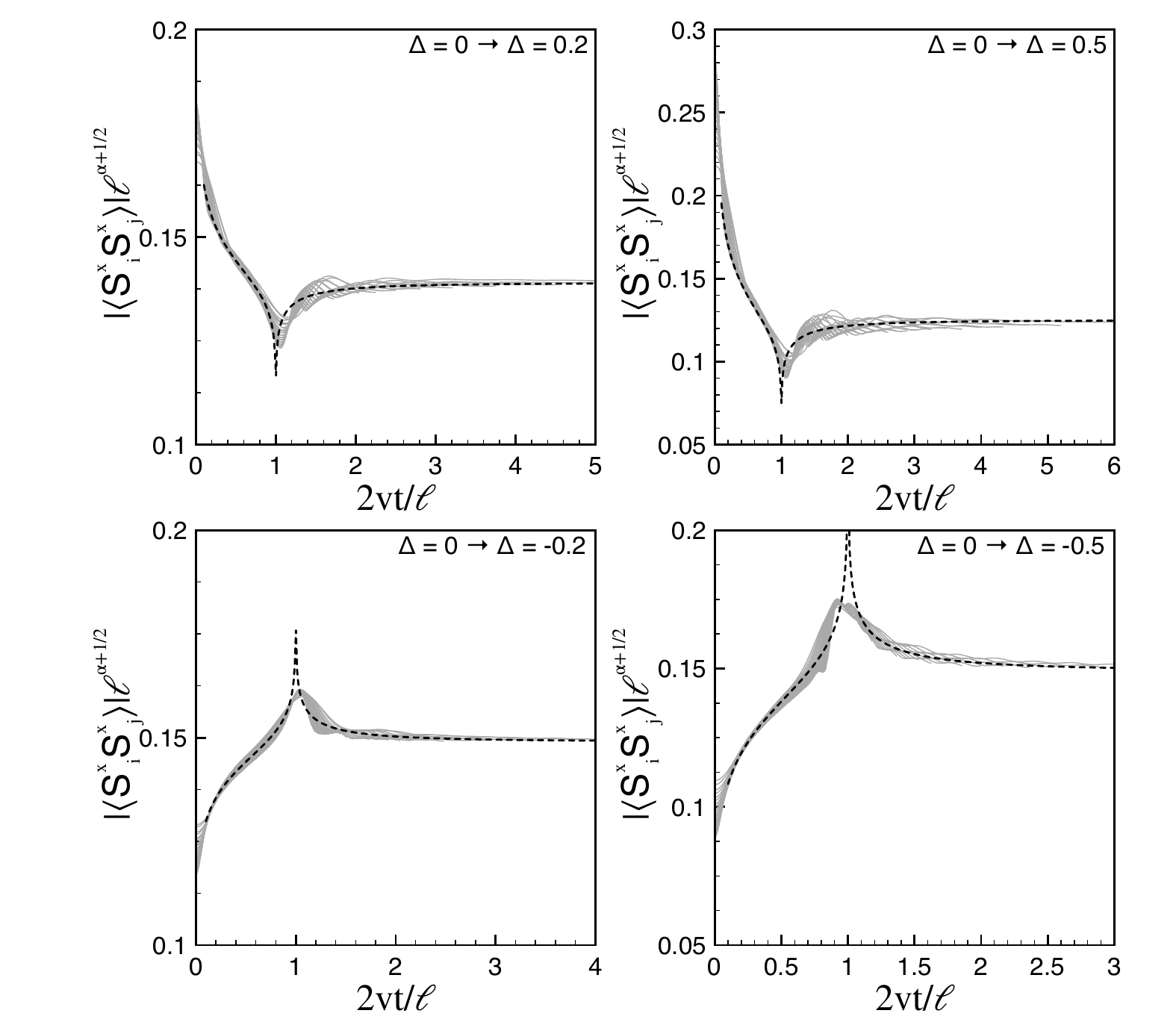}
\caption{Scaled transverse spin correlations ($\langle S^x_{i}(t) S^x_{j}(t)\rangle$) 
for a sudden quench of the anisotropy ($=$ interaction)  in the Heisenberg-Ising (XXZ) spin chain obtained using the TEBD algorithm~\cite{tddmrg} in Ref.~\cite{collura2015}. The correlations are scaled by a factor of $\ell^{\alpha+1/2}$, where  
$\ell = i - j$ and plotted vs. $2v t/\ell$ ($v$ is the spinwave velocity and $t >0$ is time); the exponent $\alpha = (K^{-2}-1)/4$, where the Luttinger parameter $K < 1$ for $\Delta > 0$ ($K > 1$ for $\Delta < 0$), is determined from the 
Bethe-ansatz solution of the XXZ model $K =\frac{1}{2}\left[1 - \frac{1}{\pi}\cos^{-1}\left(\Delta\right) \right]^{-1}$ ~\cite{cazalilla_rmp,giamarchi_book,bosonization_new,collura2015}. The Luttinger model prediction (dashed line)  (cf. Eqs.~\ref{eq:ctheta} and \ref{eq:sxcorr}) is $ |\langle S^x_{i}(t) S^x_{j}(t)\rangle| \ell^{\alpha+1/2} \sim t^{-\alpha} \left[1 - (2 v t/\ell)^2 \right]^{\alpha}$.
  In addition to the scaling predicted by the LM, 
the plot  also shows the the light-cone effect  appearing as a cusp (rounded off by short-distance lattice effects) at $2v t/\ell \simeq 1$.}
\label{fig:colluracalabrese1}
\end{figure}
In addition,  KRSM also obtained the evolution of the kinetic energy per unit length, i.e.
\begin{equation}
e_{\mathrm{imp}}(t) = -\frac{1}{2L} \sum_{m=1}^{L}  \langle \Phi_0 | e^{iHt}\left( c^{\dag}_{m+1}c_{m} + c^{\dag}_{m}c_{m+1}\right) e^{-iHt}| \Phi_0 \rangle, 
\end{equation}
where $|\Phi_0\rangle$ is the ground state of the non-interacting model (i.e. Eq.~\ref{eq:modeldd2} with $\Delta = \Delta_2 = 0$). The time-derivative kinetic energy exhibits a universal power-law decay $\sim c(K,v)/ t^{3}$ whose prefactor
$c(K,v) =  (K^2 + K^{-2}-2)v_F /16\pi v^2$~\cite{karraschmedem2012,nessi2013} is a function of the Luttinger parameter  $K$ and the Tomonaga-boson velocity, $v$~\footnote{For the Bethe-ansatz solvable XXZ, analytical expressions are known for $v$ and $K$ at half-filling~\cite{cazalilla_rmp,bosonization_new,giamarchi_book}. For $\Delta_2\neq 0$, these parameters were obtained numerically by Karrasch et al.~\cite{karraschmedem2012}.}, which  provides a further test for the universality of the LM predictions. 

\begin{figure}
\centering
\includegraphics[scale=1.0]{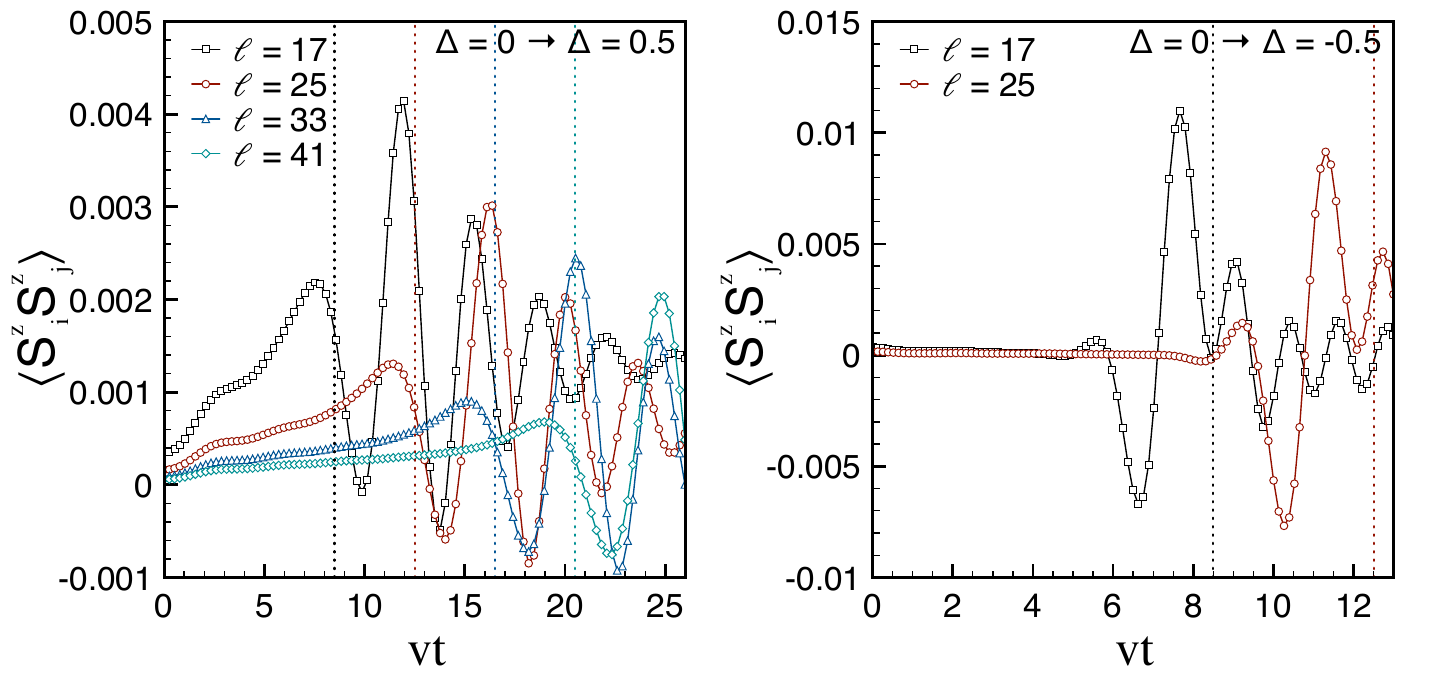}
\caption{Longitudinal spin correlation function $\langle S^z_{i} S^z_{j}\rangle$ as function of $v t$ for several separations $\ell = j − i$,  for a sudden quench of the anisotropy ($=$ interaction) in the Heisenberg-Ising (XXZ) spin chain computed numerically using the TEBD algorithm~\cite{tddmrg} in Ref.~\cite{collura2015}. As the light-cone is approached (corresponding to the vertical dashed lines for different values of $\ell$)  $t \gg |x|/v = \ell/v$, namely $\ell^{-\alpha-1}$. The oscillations are not
predicted by the Luttinger model. }
\label{fig:colluracalabrese2}
\end{figure}

 More recently, a thorough study of the universality (and lack thereof) of the LM  predictions has been undertaken by Collura, Calabrese, and Essler (CCE)~\cite{collura2015}. These authors carried out a numerical  study of a suddent quench in XXZ spin chain (cf. Eq.~\ref{eq:modeldd2} with $\Delta_2 =0$, see also Eq.~\eqref{eq:xxzmodel} below for the model in terms of spins). CCE analyzed quenches starting from the ground state with the XX model ($\Delta = 0$)  and considered several values for the Ising coupling in the post-quench Hamiltonian $\Delta = \pm 0.2, \pm 0.5$.  Using the time-evolving block decimation (TEBD) algorithm~\cite{tddmrg}, they numerically obtained the transverse and longitudinal spin-spin correlations, for which (to leading order) the LM predictions are:
\begin{align}
\langle S^x_{n+\ell}(t) S^x_n(t) \rangle &\simeq  (-1)^{|\ell|}\:  C_{V^{1}_\theta}(x = \ell a_0,t),\label{eq:sxcorr}\\
\langle S^z_{n+\ell}(t) S^z_n(t) \rangle &\simeq  2 C_{\rho_R}(x=\ell a_0,t) + 
(-1)^{|\ell|} C_{V^{2}_{\phi}}(x=\ell a_0,t).
\end{align}
where $a_0$ is the lattice parameter, and  $C_{V^{m=1}_{\theta}}(x,t)$, $C_{\rho_R}(x,t)$ and $C_{V^{2m=2}_{\phi}}(x,t)$ are given by Eqs.~\eqref{eq:ctheta},
\eqref{eq:cphi}, and \eqref{eq:rhor}, respectively. CCE found a fairly good agreement of the LM predictions with the numerics  for the transverse spin correlations, $\langle S^x_{i+l}(t) S^x_i(t) \rangle $ (see Fig.~\ref{fig:colluracalabrese1}). However,  the agreement with the LM predictions for the longitudinal correlations, $\langle S^z_{i+l}(t) S^z_i(t) \rangle$, was found to be much poorer (see Fig.~\ref{fig:colluracalabrese2}). CCE convincingly argued that the lack of agreement in the latter case stems from the different character of the  $S^z_n$ spin operators, as compared to $S^x_n$. Indeed,  
the spin operator measuring the projection on the $z$-axis reads:~\cite{bosonization_new,giamarchi_book}: 
\begin{equation}
S^z_n = c^{\dag}_n c_{n} - \frac{1}{2} \simeq   \frac{1}{\pi} \partial_x \phi(x=n a_0) + (-1)^n  \cos 2\phi(x = n a_0) + \cdots
\end{equation}
that is, a rather local operator in terms of the Jordan-Wigner  fermion operators $c_n$ and $c^{\dag}_n$~\cite{giamarchi_book,cazalilla_rmp}. On the other hand, the spin operator along the $x$ axis, 
\begin{equation}
S^{x}_n = (c_{n} + c^{\dag}_n) \prod_{l<n} \left(1-2c^{\dag}_l c_l\right) = (-1)^n \cos \theta(x = n a_0) + \cdots, \label{eq:sx}
\end{equation}
is  rather non-local operator due to the Jordan-Wigner string $\prod_{l<n} \left(1-2c^{\dag}_l c_l\right) $~\cite{giamarchi_book,cazalilla_rmp}. To fully appreciate the impact of this difference, let us recall that the (initial) state, which is the ground state of the  XX model (i.e. $\Delta = 0$),  can  be written as a non-interacting Fermi sea of the Jordan-Wigner fermions, i.e. $|0\rangle = \prod_{k < \frac{\pi}{2}} c^{\dag}_k | \mathrm{vac}\rangle$, where $|\mathrm{vac}\rangle$ is the Fermion vacuum state and $c_k = L^{-1} \sum_{n} e^{-i k n} c_n$. 
\subsection{Pre-thermalization and quench in a 2D Fermi liquid}\label{sec:pretherm}
 
Pre-thermalization was  discussed by
Berges and coworkers in the context of high-energy ion collisions~\cite{berges}. It refers to metastable state of a system that has been driven out of equilibrium rapidly establishes a kinetic temperature based on the average kinetic energy. Despite this fact,  the eigenmode distribution of the in the metastable state does not correspond to a Bose-Einstein (for bosons) or a Fermi-Dirac (for fermions) distribution. These ideas have found resonance in the
study of non-equilibrium (quench) dynamics of ultracold atomic systems.  Moeckel and Kehrein~\cite{moeckel2008} discussed them in relation to a two-stage thermalization scenario that should take place following a quantum quench.
In their work, Moeckel and Kehrein studied a quench of the interaction in the Hubbard model using the flow equation method~\cite{moeckel2008}. Considering the infinite-dimensional Hubbard model on the Bethe lattice, they  obtained the short to intermediate time evolution of the momentum distribution. Their result shows some striking resemblance with the results described earlier for the LM~(cf. Fig.~\ref{fig:fpt}). However, one major difference with the LM  is that the discontinuity at the Fermi surface does not close completely. Instead, Moeckel and Keherein found that it saturates at a constant value $Z_{\mathrm{neq}}$ which obeys the relation $(1-Z_{\mathrm{neq}}) = 2\left( 1 - \mathrm{Z}_{\mathrm{eq}}\right)$, where $Z_{\mathrm{eq}} < 1$ is the discontinuity in the momentum distribution in the  \emph{interacting} ground state.  

 Pre-thermalization has also been discussed in connection to study of quantum quenches in ultracold bosonic gases in Schmiedmayer's group in Viena~ \cite{pretherm1,pretherm2}. However, in this section, we focus on reviewing the results for the pre-thermalization dynamics for a two-dimensional (2D) gas of spinless fermions interacting with long-ranged interactions, which, in a certain sense,  can be regarded as the 2D generalization of the LM. The dynamics ensuing a quench of the interaction in this system has been studied by Nessi, Iucci, and Cazalilla (NIC)~\cite{nessi2014}, and it may be of relevance for the study of the non-equilibrium dynamics of dipolar quantum gases~\cite{dipolar_fermigases}. The Hamiltonian of the model studied by NIC reads:
\begin{align}
H &= H_0 + H_{\mathrm{int}},\label{eq:2dfl} \\
H_0 &= \sum_{k} \epsilon(\vec{k}) \psi^{\dag}(\vec{k}) \psi(\vec{k}),\\
H_{\mathrm{int}} &= \frac{1}{2\Omega} \sum_{\vec{k}\vec{p}\vec{q}} V(\vec{q})
\psi^{\dag}(\vec{k}+\vec{q})\psi^{\dag}(\vec{p}-\vec{q}) \psi(\vec{p}) \psi(\vec{q}),
\end{align}
where $\psi(\vec{k})$ ($\psi^{\dag}(\vec{k})$) annihilate (create) fermions
with momentum $\vec{k}$, where $\vec{k} = (k_x, k_y)$ is a two dimensional vector. As in the case of the LM, it was assumed by NIC that the system is prepared in the ground state of the non-interacting Hamiltonian $H_0$ and it evolves according to the interacting Hamiltonian $H$ for $t \geq 0$, which is tantamount to a quench of the interaction. The Fourier transform of interaction potential $V(\vec{q})$ is assumed to vanish rapidly for $|\vec{q}|\gg q_c$, where
$q_c \sim R \gg p^{-1}_F$, $R$ being the range of the interaction and $p_F = \sqrt{4\pi \rho_0}$ the Fermi momentum ($\rho_0$ is the areal particle density).  In order to render the expressions analytically tractable, the Fourier transform of the interaction potentila is taken of to be of form $V(\vec{q}) = f_0\, (q/q_c)^n e^{-q/q_c}$, where $f_0$ parametrizes
the interaction strength and $n = 0, 1, \ldots$ is a positive or zero integer (see below). 
\begin{figure}
\centering
\includegraphics[scale=1.5]{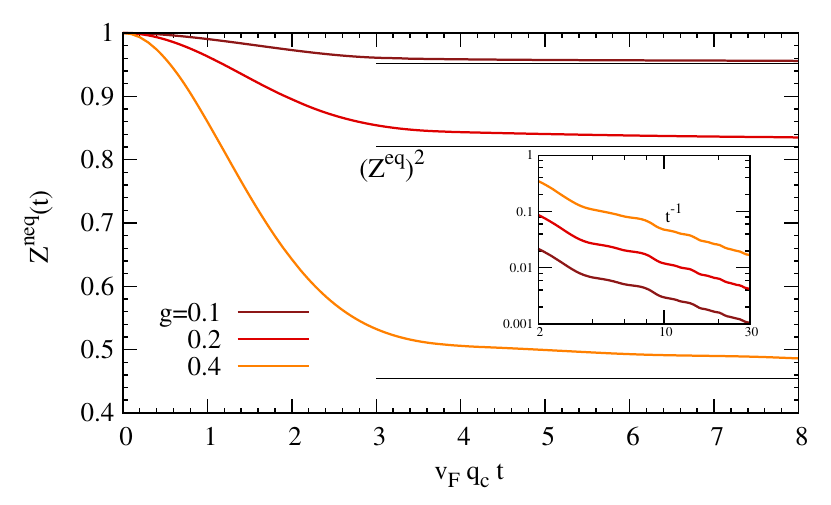}
\caption{Time-evolution of the discontinuity at the Fermi momentum following
an quench of the interaction in a two-dimensional Fermi liquid with long
range interactions~\cite{nessi2014}. The short time behavior is $Z(t) \simeq 1 - c t^2$, where $c \propto f^2_0$, where $f_0$ is the strength of the interaction. The inset shows better detail of the way  $Z(t)$ approaches its constant asymptote as $t^{-1}$. The dimensionless inteaction coupling $g = f_0\sqrt{p_F q_c/2\pi^2 v_F}$.
The horizontal lines indicate the long-time pre-thermal asymptote, $Z_{\mathrm{neq}}=(Z_{\mathrm{eq}})^2$, which is obtained from the Fermi-surface bosonization treatment (cf. section~\ref{sec:pretherm}).}
\label{fig:ztfl}
\end{figure}

To access the short to intermediate time dynamics of the model, NIC first carried out a perturbative analysis to leading (i.e. second) order in the interaction. Thus, they showed that, the discontinuity at $p_F$ of the momentum distribution, $Z(t)$, also exhibits a plateau similar to the one observed by Moeckel and Kehrein~\cite{moeckel2008} in the  case of the Hubbard model (see also Eckstein \emph{et al.}~\cite{eckstein2009}). This plateau indicates the existence of a pre-thermalized state. Furthermore, NIC found that, to the leading order in $f_0$,  the relationship  $1- Z_{\mathrm{neq}} = 2(1-Z_{\mathrm{eq}})$
also holds for the 2D Fermi gas described by the above model.  
 
In addition, in order to understand the emergence of the pre-thermalized state, NIC resorted to the Fermi surface (FS) bosonization  method~\cite{FSbosonization}. This method has been applied in equilibrium and it provides a non-perturbative foundation to  Landau's Fermi liquid theory. Unlike the equilibrium case where the method is applied to a low-energy effective Hamiltonian~\cite{FSbosonization}, NIC applied it to the bare Hamiltonian, $H$ (cf. Eq.~\ref{eq:2dfl}), which describes the interactions between the bare fermions. Performing a truncation of $H$ that amounts to neglecting inelastic scattering between the fermions and keeps only forward and exchange interactions,  NIC~\cite{nessi2014}  
re-wrote $H$ in terms of the Fourier components of the density operator~\cite{FSbosonization}:
\begin{equation}
H\simeq \sum_{\vec{S},\vec{T}} \sum_{|\vec{q}|\ll \lambda} \rho_{\vec{S}}(\vec{q})
\left[\frac{v_F}{\Omega} + \frac{V(\vec{q})}{A} \right] \rho_{\vec{T}}(-\vec{q}),\label{eq:bos2dfl}
\end{equation}
where $v_F = |\vec{\nabla}_{\vec{k}}
\epsilon(\vec{k})|$ is the Fermi velocity and $A$ is the area of the system. In writing Eq.~\eqref{eq:bos2dfl}, it is assumed that a crown  of width $\lambda$ around the FS has been  ``sectorized''  into $N$ squat patches of transverse size $\Lambda$, such that 
$p_F \gg \Lambda \gg \lambda > q_c$~\footnote{The number of patches, $N$, must be taken to be large but finite, in order to keep under control the divergences in the Cooper channel leading to the Kohn-Luttinger instabilities~\cite{rg_shankar}.}. The constant $\Omega = \Lambda A/(2\pi)^2$. As in the case of TLLs, the Fourier components of the density, $\rho_{\vec{S}}(\vec{q})$ obey a KM algebra~\cite{FSbosonization} ($|\vec{q}|,|\vec{q}^{\prime}|\ll \lambda$, compare with Eq.~\ref{eq:km}):
\begin{equation}
\left[ \rho_{\vec{S}}(\vec{q}), \rho_{\vec{T}}(\vec{q^{\prime}}) \right] = \Omega \left( \vec{q} \cdot \vec{\hat{n}}_{\vec{S}} \right) \delta_{\vec{S},\vec{T}}\delta_{\vec{q}+\vec{q^{\prime}},\vec{0}}, \label{eq:hambos}
\end{equation}
where $\vec{\hat{n}}_{\vec{S}}$ is a unit vector normal 
to the circular FS at the patch position of $\vec{S}$.
This  equation turns the  diagonalization of  
Eq.~\eqref{eq:bos2dfl} into a problem akin
to a system of (chiral) Luttinger models (one for
each FS patch) coupled by forward-scattering 
interaction. Like in the case of the LM, the KM
algebra allows us to obtain a diagonal representation 
of Eq.~\ref{eq:hambos} in terms of a set of 
bosonic eigenmode operators:
\begin{equation}
H \simeq \sum_{l,\vec{q}} \omega_l(q) b^{\dag}_{l}(\vec{q})b_l(\vec{q}).\label{eq:exactsol}
\end{equation}
This expressions clearly shows that
the short to intermediate-time dynamics described by $H$ can be approximated by a exactly solvable Hamiltonian,
whose dynamics is strongly constrained by the integrals
of motion $I_l(\vec{q}) = b_l(\vec{q}) b_l(\vec{q})$ (more 
on this further below).

In addition, the diagonalization of $H$ (neglecting 
inelastic scattering between fermions) allowed NIC to
obtain non-perturbative results for the evolution of the
post-quench single-particle density matrix. Using similar expressions to the bosonization formula, Eq.~\eqref{eq:bosonization}~\cite{FSbosonization}, the following results for the behavior of the discontinuity in the momentum distribution at the FS were obtained~\cite{nessi2014}:
\begin{align}
Z(t \ll R/v_F) &= \exp \left[ - c t^2  \right],\\
Z(t \gg R/v_F) &= Z_{\mathrm{neq}} e^{g^2 a_n (v_F q_c t)^{-(2n+1)}}.\label{eq:longt}
\end{align}
where $c$ depends on the details of the interaction~\cite{nessi2014} and $g = f_0\sqrt{p_F q_c/2\pi^2 v_F}$.
The long-time asymptote of $Z(t)$ obeys $Z_{\mathrm{neq}} =  (Z_{\mathrm{eq}})^2$, which is the non-perturbative version of Moeckel and Kehrein's relationship between the equilibrium and pre-thermalized values of the discontinuity at $p_F$~\cite{moeckel2008}.  The full crossover from the short time limit (which agrees with the perturbative results~\cite{nessi2014}) to the long time limit of Eq.~\eqref{eq:longt} was obtained by evaluating the integrals numerically and it is shown in Fig.~\ref{fig:ztfl}.

 Thus, the explicit construction of the exactly solvable truncation of $H$ gives access to a non-perturbative solution  of the short to intermediate time dynamics (up to times $t \gtrsim f^2_0 N(0) $ ($N(0)=k_F/2\pi v_F$ being the density of states at the Fermi level), at which inelastic collisions kick in and should drive the system to a thermal state. In addition, it also clarifies the physical  origin of the phenomenon of pre-thermalization by relating it to constrained dynamics of the exactly solvable model of Eq.~\eqref{eq:exactsol}, which is as we will exactly in 
section~\ref{sec:entgge} relaxes not to the  grand canonical ensemble, but to the generalized Gibbs ensemble due to the combination of  Gaussian nature of the initial state and the dephasing between the bosonic FS eigenmodes of \eqref{eq:exactsol}.  

\section{Smooth quantum quenches}

\subsection{Analytical results}

In the previous section, we have focused  on the dynamics following a sudden quench of the interaction. However, both from the theoretical and experimental point of view,  it is interesting to consider the question of how the system may evolve under a smooth quench. In the case of the LM, it  allows to  study the non-equilibrium dynamics of the system as a function of the rate with which the interaction is turned on, sudden quenches corresponding to the the fastest rate and  the adiabatic limit to the slowest one.  Therefore, the study of the smooth quenches can be used to compare the effects of a sudden quench to an adiabatic evolution and to better understand the mechanisms by  which quantum many-body  systems are driven out of equilibrium. In addition, time scale which controls the change in the Hamiltonian parameters the non-equilibrium dynamics with the question
how it compares to other characteristic time scales of the problem is an important input for experiments studying quench dynamics.  

 Another interesting question that can be addressed in this context is how much the time-evolved state of the system is reminiscent of the original starting state. In this regard, the Loschmidt echo, or fidelity in the quantum
information-theoretic language, measures the overlap between the state of the smoothly quenched system and the initial state. This measurement provides direct
insight into the many-body dynamics and  the relation
between quench dynamics and quantum information.

For a smooth-quench the Hamiltonian becomes explicitly time-dependent:
\begin{equation}
    H(t) = \sum_{q\neq 0} \left\{ \omega(q,t) a^{\dagger}(q) a(q)  +
    \frac{1}{2} g_2(q,t) |q| \left[ a(q) a(-q) + a^{\dagger}(q) a^{\dagger}(-q) \right] \right\}.  \label{eq:slowqham}
\end{equation}
where $\omega(q,t) = \left[ v_F + g_4(q,t) \right]|q|$.  The time evolution is described by the Heisenberg equation of motion:
\begin{equation}
   i \partial_t a(q,t) = [a(q,t), H(t)] = \omega(q,t) a(q,t)
   + g_2(q,t) |q| a^{\dagger}(-q,t), 
\end{equation}
and similarly
  \begin{equation}
   i \partial_t a^{\dagger}(-q,t) = - \omega(q,t)
   a^{\dagger}(-q,t)  - g_2(q,t) |q| a(q,t).  
\end{equation}
This equation of motion can be solved by a similar ansatz to the one used for the sudden quench:
\begin{equation}
     a(q,t) = f(q,t) a(q) + g^{\star}(q,t) a^{\dagger}(-q).
\end{equation} 
However, in this  case the functions $f(q,t)$ and $f(q,t)$ obey the Bogoliubov-deGennes (BdG) equations of motion:
\begin{equation} \label{eqn:BogoUV}
    i \partial_t \left[ \begin{matrix} f(q,t) \\
        g(q,t)  \end{matrix} \right] =  \left[ \begin{matrix}
        \omega(q,t) & g_2(q,t)|q| \\ -g_2(q,t)|q| & -\omega(q,t) 
          \end{matrix} \right]   \left[ \begin{matrix} f(q,t) \\
        g(q,t)  \end{matrix} \right],
 \end{equation}
which are supplemented by the initial conditions: $f(q,0) = 1$ , and $g(q,0) =0$ and the constraint $|f(q,t)|^2 - |g(q,t)|^2=1$, which is required by Bose-Einstein statistics.  Thus,  $f(q,t), g(q,t)$ contain all the dynamical information about the smooth quench and, as in the case of the sudden quenches,  expectation values of the time-dependent observables can be obtained from their knowledge. 

 Dora, Haque and Zar{\'a}nd (DHZ) \cite{DoraHaqueZarand} obtained solutions to the BdG equations for a linear ramp of the the interaction assuming that $\omega(q,t) = \omega(q) = v_0(q) |q|$ is independent of time and 
\begin{equation}
g_2(q,t) = g_2(q)  Q(t)
 \end{equation}
where   $Q(t) = t \Theta(t(t-\tau))/\tau + \Theta(t-\tau)$, with
$\Theta(t)$ the Heaviside function. Note that $Q(t>\tau) = 1$
and $Q(t<0) =0$, that is, $\tau$ determines the characteristic quench time. For $\tau \rightarrow 0$, a sudden quench is obtained, while the adiabatic
limit is approached by letting $\tau \rightarrow \infty$. Assuming a 
perturbatively small $g_2(q) = g_2 e^{- |q| R/2}$, DHZ obtained the
following asymptotic form for the
instantaneous single-particle density matrix (cf. Eq.~\ref{eq:opd}) for 
$t \gg \tau$:
\begin{equation} \label{eqn:SQtoAD}
   \frac{C_{\psi_R}(x,t)}{C_{\psi_R}^{(0)}(x)} \sim
   \left\{ \begin{matrix}  A\left(\frac{\tau}{\tau_0}\right)\left(\frac{R}{\min{(|x|,2v_0t})}
       \right)^{\gamma_{SQ}}  &  \mbox{for} \, |x| \gg 2 v_0 \tau \\ \left(
         \frac{R}{|x|}\right) ^{\gamma_{ad}}  &  \mbox{for}\,  |x| \ll 2
       v_0 \tau \end{matrix} \right.
\end{equation}
where $\gamma_{SQ} = g_2^2(q=0)/v_0^2+ O(g^3_2)$ and $\gamma_{ad} =
g_2^2/2v_0^2 + O(g^3_2)$ ($v_0 = v_0(q = 0)$) are the (perturbative) sudden quench and
adiabatic-limit exponents. The prefactor $A(\tau/\tau_0)$ depends on the
speed of the quench: For a sudden quench, $A(\tau/\tau_0\ll 1)
\sim 1$, while for smooth quench  $A(\tau/\tau_0 > 1)
\sim  (\tau/\tau_0)^{\gamma_{ad}}$. 
The physical explanation for two kinds of behavior displayed in Eq.~\eqref{eqn:SQtoAD} lies in the following crossover behavior:  When the interaction is quenched at a rate $\sim \tau^{-1}$, the ``slow'' excitations of
energy $\omega(q) < \tau^{-1}$ experience it as a sudden quench. 
On the other hand, fast excitations with energies $\omega(q) > \tau^{-1}$ can adjust the the change of the interaction strength adiabatically. Since high (low) energy
excitations determine the short (long) distance correlations,
the tail of $C_{\psi_R}(x,t)$ is governed by the sudden quench exponent, whilst
its short distance behavior is described by the adiabatic exponent.

 The subject of smooth quenches has often been related to the dynamics across critical points (see e.g.~\cite{KZMReview} and references therein). However, for the quench of interaction in LM only one phase (critical) is involved. For example, for the XXZ model, only the critical region is related to the LM, and for Bose Hubbard models, LM corresponds to  the superfluid regime.  Therefore, the Kibble Zurek mechanism, which is related to the production of  topological  defects  when quenching a system across a critical point,  is not relevant for  smooth quenches of the interaction in the LM. Nevertheless, it is possible to discuss the  production of  quasi-particles in a smooth quench of the interaction, as it was done by Dziarmaga and Tylukti~\cite{Dziarmaga}.  The average density of excited quasiparticles $n_{ex}$ scales with $\tau$ as $n_{ex} \sim \tau^{-1}$, while the more directly measurable excitation energy density scales like $\varepsilon \sim \tau^{-2}$ at zero temperature. On the other hand, at finite temperature $\varepsilon \sim \tau^{-1}$. At zero temperature,  Dziarmaga and Tylukti also showed that the production of excitations does not change the algebraic $x^{-2}$ decay of the  density (cf. Eq.~\ref{eq:denscor}). Instead (relative to the  initial state), they only yield an additive correction to the prefactor. This behavior contrast
with the exponential decay of  the correlations induced by the Kibble-Zurek mechanism following a quench from a disorder to an ordered phase.
\subsection{Comparison to numerical approaches}
As in the case of sudden quenches, the LM predictions for the  correlation functions and other quantities (see below)  in the case of a smooth quench have been compared to numerical calculations.  In this section, we review some of the most important results in this regard.

\begin{figure}
\centering
\includegraphics[scale=1.5]{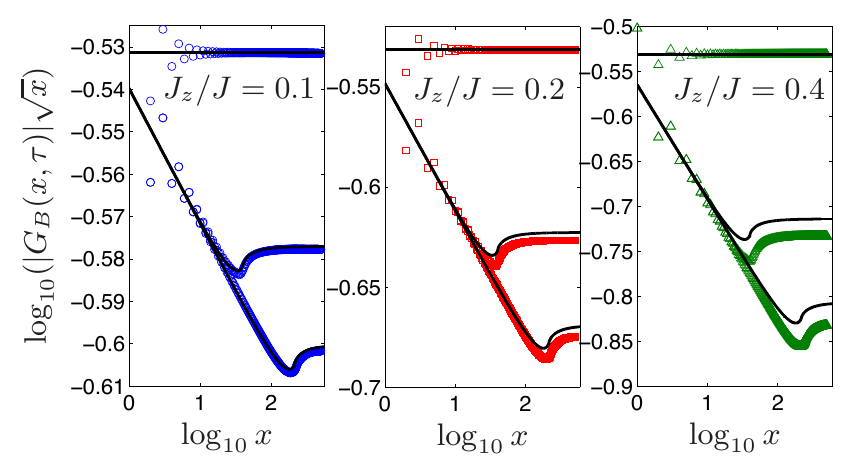}
\caption{Post-quench spin-flip correlation function for $J_z/J = 0.1$ (left   panel), $0.2$ (middle panel), and $0.4$ (right panel) for  $J\tau=0$, $20.2$ and $108.3$ (from top to bottom) compared with the Luttinger model prediction of Eq.~\eqref{eqn:GB}. At short distances, the numerical results are influenced by lattice effects, and therefore deviate from the Luttiger model predictions.}
\label{fig:PollmannDora}
\end{figure}

In order to test the correctness of the LM predictions for smooth quench dynamics, Pollmann, Haque and D{\'o}ra (PHD) numerically studied the anisotropic (XXZ) Heisenberg model  in the critical regime by using  the infinite time-evolving block decimation algorithm (iTEDB)~\cite{tddmrg}. They considered the Hamiltonian:
\begin{equation}
     H_{XXZ} = \sum_{m} \left[ \left(S_{m}^{x} S_{m+1}^{x} + S_{m}^{y}
       S_{m+1}^{y}\right) + J_z(t) S_{m}^{z} S_{m+1}^{z}
       \right], 
 \label{eq:xxzmodel}      
\end{equation}   
assuming antiferromagnetic exchange interaction, i.e. $J>0$ and quenching the Ising exchange,  $J_z(t)$  according to $J_z(t) = J_z Q(t)$ with $Q(t<\tau) = t/\tau$ and $Q(t > \tau) =1$. As explained in section~\ref{sec:quest}, using  the Jordan Wigner transformation (see e.g. Refs.~\cite{bosonization_new,giamarchi_book,cazalilla_rmp}) this model can be mapped to an interacting lattice model of spinless  fermions in 1D (cf. Eq.~\ref{eq:modeldd2} with $\Delta_2 = 0$). Bosonization~\cite{bosonization_new,giamarchi_book} then allows to relate this model to a smooth quench in the LM  (cf. \ref{eq:slowqham})  with
   $\omega(q) = v|q|$ ($v=J$) and $g_2(q,t) = g_2(q) Q(t)$, $g_2(q) = g_2    e^{-|q| R/2}$, being $R\sim a_0$ a short-distance cut-off of the order of the lattice parameter $a_0$. $R$ was numerically obtained by PHD to be $R =0.5622 a_0$. In addition, the  value of $g_2$ is fixed using perturbation theory, which requires that $-1 \ll g_2/2J = J_z/\pi J   \ll 1$. 

The staggered part of the transverse magnetization is given by Eq.~\eqref{eq:sx}. Hence,  in the scaling limit where $|x|, v\tau \gg R$ after the quench at $t=\tau$, the transverse post-quench correlations can be evaluated using bosonization to yield:
\begin{equation} \label{eqn:GB}
    C_{S^{+}}(x,\tau) \approx \frac{C(-1)^{|x|}}{\sqrt{|x|}}
    \left|\frac{R}{x} \right|^{\frac{g_2}{2v}}
    \exp{\left[-\frac{g_2}{2v} f\left(\frac{x}{2v\tau}\right) \right]} 
\end{equation}
where $f(y) =1/2\sum_{s=\pm1} s(y-s) \log|y-s|$, and $
C = 2^{-1/3}e^{1/2} A^{-6}$, where $A=1.28243\dots$  is Glaisher's constant. In Fig.\ref{fig:PollmannDora}, the 
comparison of this  analytical result with the numerics from iTEDB is shown. The agreement is excellent for small $J_z/J$, although it worsens slightly for higher values of $J_z$.

In addition to correlation functions,  Dora, Pollmann, Fort{\'a}gh and Zar{\'a}nd (DPFZ) have studied
the Loschmidt echo (LE) \cite{LE} as a many-body generalization of the orthogonality catastrophe~\cite{anderson_oc}. The LE is 
defined as the overlap of two wave functions
$|\Psi_0(t)\rangle$ and $|\Psi (t)\rangle$ evolved from the same
initial state $|\Psi_0\rangle$ but with different Hamiltonian ($H_0$ and $H$): 
\begin{equation} 
 {\cal L}(t) =\mid \langle \Psi_0 \mid e^{i H_0 t } e^{-iH
      t} \mid \Psi_0\rangle \mid^2.
\end{equation}
In quantum information theory this quantity is also called fidelity , and it is a measure of the distance between two quantum states which can be used to identify irreversibility and chaos. Moreover, like entanglement measures, this quantity can be used to detect quantum phase transitions (see e.g.~\cite{review_polkovnikov,fidelity_qpt} and references therein). 

In their paper on LE, DPFZ considered a quench between two different values of the interaction, which corresponds to setting $g_2(q,t) = g_i(q) + \Delta g(q,t)$
in Eq.~\eqref{eq:slowqham}. Here $g_i(q)$ is the initial value of the interaction, $\Delta g(q,t) = [g_f(q) - g_i(q)] Q(t)$ (where  $Q(t)$ has been defined above for a linear ramp of the interaction), and $g_f(q)$ is the final value of the interaction.  The initial and final quasi-particle spectra are given by $\omega_{i(f)}(q) = \sqrt{\omega^2(q) - g_{i(f)}^2(q)q^2]}$,
and the Luttinger parameters are characterized by those interaction
strengths in the following way:
\begin{equation}
    K_{i(f)} = \sqrt{ \frac{\omega(q) - g_{i(f)}(q) |q| } {\omega(q) +
          g_{i(f)}(q) |q| } }. 
\end{equation}
Thus, DPFZ  expressed  the LE analytically in terms of the Bogoliubov coefficients of Eq.~\eqref{eqn:BogoUV}):
\begin{equation}
  {\cal L}(t) = \exp{\left(-\sum_{q>0} \log{\left[ \left|f(q,t)\right|^2 \right]}  \right)},
\end{equation}
which expresses the LE in terms of the number of excited quasi-particles 
in the final state.  Using  $|f^{ad}(q,t)|^2 = 1/2+(K_i/K_f + K_f/K_i)/4 $ for
 $t> \tau$, DHPZ obtained for the the LE in the adiabatic limit
the following expression~\cite{fidelity}:
\begin{equation}
    {\cal L}_{ad} = \left[ \frac{1}{2} + \frac{1}{4}
      \left(\frac{K_i}{K_f} + \frac{K_f}{K_i} \right)\right]^{-\frac{L}{\pi a_0}},
\end{equation}
were $L$ is the system size and  and $a_0$ is the short-distance cut-off. Therefore, the LE decays exponentially with the system size, $L$.

 On the other hand, for a sudden quench the long time limit of the LE
takes a different form:
\begin{equation}
{\cal L}_{SQ}\left(t \gg \frac{a_0}{2v_0}\right) = {\cal L}_{ad}^2. 
\end{equation}
This can be understood in the following way: The LE of an adiabatic quench
involves only the ground states of the initial and final Hamiltonian, i.e., square of the  overlap of the ground states $\langle G |
G_0\rangle$ where $|G_0\rangle$($|G\rangle$) is the ground state of $H_0$($H$). For a sudden quench, inserting the resolution of the identity operator in terms the eigenstates of the post quench Hamiltonian $H$ between the two evolution operators in $\langle \Psi_0 |e^{i H t} e^{-i H_0
  t} |\Psi_0\rangle$, and after taking into account the  dephasing of the energy phase factors of different excited states the ground state contribution remains. Asymptotically in the
thermodynamic limits $|\langle \Psi_0 |e^{i H t} e^{-i H_0
  t} |\Psi_0\rangle| = |\langle G |
G_0\rangle|^2$. Hence, ${\cal L}_{SQ} = |\langle \Psi_0 |e^{i H t} e^{-i H_0
  t} |\Psi_0\rangle|^2 = |\langle G |G_0\rangle|^4 = {\cal L}_{ad}^2$
for $t \rightarrow \infty$.  These results for the LE were  numerically confirmed by DPFZ for the XXZ spin-chain model using matrix-product states~\cite{DMRGBook}. 

 Considering on a smooth quench of the interaction in a 1D Bose gas which can be described as a TLL~\cite{cazalilla_rmp}, Bernier and coworkers addressed the  question how  the system is driven  out of equilibrium by an increasing quench rate~\cite{Orignac}. Working in the weakly interacting limit, where $v(t) K(t)$ remains time-independent and $u(t)/K(t) = (v_0/K_0) (1 + \tau/\tau_0)$
 ($\tau_0 = \pi v_0 t\tau_1/K_0(g_f-g_i)$),
they obtained solutions of the equations of motion for
the Fourier components of the density $\phi(x,t)$ and
phase $\theta(x,t)$ fields in terms of Bessel functions.  
This allowed them to evaluate the phase correlation functions of the bosons,
$\langle e^{i\theta(x,t)} e^{-i\theta(0,t)} \rangle$,
which they compared with the numerical 
results obtained using td-DMRG~\cite{tddmrg} for the Bose-Hubbard model~\cite{cazalilla_rmp}. Similarly to Dora \emph{et al.}, they found that, at short distances,
the exponent determining the decay of the correlations with the distance is given by the results for adiabatic quench, while for long distances, correlations decay with a power-law exponent approaching the sudden quench value. The long distance regime is separated from an intermediate distance using a generalized Lieb-Robinson bound~\cite{liebrobinson}, that is, a length scale defined as $\xi_B = 2/l_0 \int_{0}^{t} dt' v(q=0,t') $ with $l_0=v_0^2 \pi \tau/[K_i (g_f-g_i)]$~\cite{Orignac}. This behavior can be regarded as a  generalization of the the light-cone-effect discussed above for the sudden quenches.  

\section{Steady state and the generalized Gibbs ensemble}\label{sec:entgge}

As described in section~\ref{subsec:prel}, one of the main motivations for the study of a quantum quench of the interaction in the LM was to find an analytically tractable  model for which it is possible to demonstrate the absence of thermalization. Indeed, it turns out that the LM provides an excellent toy model to understand this phenomenon. As we discuss in this section, it also provides an excellent playground to understand the emergence of the Generalized Gibbs ensemble  (GGE) as an effective description of the correlations in the steady state ofollowing the quantum quench. 

In order to gain a broader perspective, let us first discuss the GGE as it was originally introduced~\cite{rigoletal2006}. Rigol and coworkers noticed that dynamics of integrable models  is strongly constrained by the presence of a large number of integrals of motion, $I(q)$ ($[H,I(q)] = 0$, where $H$ is the post-quench Hamiltonian). Thus, relying on Jaynes's fundational work on statistical mechanics, they proposed that the steady steady state of integrable models following a quantum quench is described by the following density matrix:
\begin{equation}
\rho_{\mathrm{GGE}} = \frac{e^{-\sum_{q}\lambda(q) I(q)}}{Z_{\mathrm{GGE}}}.
\end{equation}
This result can be obtained by maximizing the von Neumann entropy $S = - \tr\left[\rho \log \rho \right]$
subject to the constraints imposed by the conservation of the $I(q)$. Note that if the set of integrals of motion includes only the total energy $H$ and particle number $N$, the resulting density matrix is the Gibbs' grand canonical ensemble~\cite{jaynes}. In such a case, the Lagrange multipliers $\lambda(q)$  correspond to the familiar inverse absolute temperature $\beta = 1/T$ and the ratio of (minus) the chemical potential to the absolute temperature, $-\mu/T$. However, in general, as it is the case of integrable systems, the set of conserved quantities is larger than $\{H, N\}$, and more $\lambda(q)$ are required. The latter are determined from the  initial conditions by requiring that
\begin{equation}
\langle I(q) \rangle_{\mathrm{GGE}} = \tr \rho_{\mathrm{GGE}} I(q) = \tr \rho_0 I(q) = \langle I(q) \rangle_0,
\end{equation}
where $\rho_0$ is the density matrix describing the initial state.  Rigol and coworkers have provided convincing numerical evidence  for the GGE by performing  a careful numerical analysis  of several types of quenches in the lattice hardcore Bose gas~\cite{rigoletal2006,rigolgge2,rigolgge3}. For this system, they identified~\cite{rigoletal2006} the set of integrals of motion $I(q)$ to be the occupation operators of the eigenmodes of the system, which  are the occupations of the Jordan-Wigner fermions in momentum space~\cite{cazalilla_rmp}. Subsequently, an analytical proof that the GGE describes the correlations in the asymptotic state of the quenched LM model was provided in Ref.~\cite{cazalilla2006}.

Yet, the effectiveness of the GGE description must be regarded as something rather non-obvious and even striking. It is striking that a density matrix corresponding to 
a mixed state can describe the result of the unitary evolution of a pure state such as the ground state of the non-interacting  LM~\footnote{Indeed, the description is effective at the level of correlation functions. However, it has been pointed out~\cite{iucci2009,chung2012} that the GGE does not contain all the necessary correlations between the eigenmodes to reproduce other quantities such as the energy fluctuations. In fact, there are situations for which the GGE essentially looks like a thermal density matrix~\cite{rigolgge2,chung2012} and therefore it appears as if the system exhibits thermal correlations. However, the difference with a real thermal state is exhibited by the failure of the system in the steady state to obey the fluctuation-dissipation theorem for the energy fluctuations~\cite{chung2012}.}. It is also not obvious that the number of integrals of motion required to construct the GGE  is only a particular subset  of all the possible integrals of motion of the model. Some answers to these questions can be extracted from the theory of quantum entanglement.  

As we have seen above, quantum entanglement plays a important role in the physics of quantum quenches. The light-cone effect described in section~\ref{sec:corrdyn}  can be traced back to the propagation of entangled pairs of quasi-particles. In addition, we will see below that the GGE emerges as a consequence of decoherence caused by the time evolution which erases all but a certain kind of correlations that exist amongst the eigenmodes in the
initial state of the system.  This observation is applicable to a certain class of initial states, known
as gaussian states, which are defined further below 
in this section. However, before discussing the connections between the GGE and entanglement, it is worth providing
a short pedagogical introduction to the most important concepts of entanglement theory, which is undertaken in the following section. 
\subsection{Entanglement, reduced density matrices, and entanglement spectra}
Entanglement is one of the most remarkable features of quantum mechanics. It was introduced by Schr{\"o}dinger when he used the German word  
``\emph{Verschr{\"a}nkung''} (translated into English as ``entanglement'') to describe
the correlations between two particles that interact and then  separate when addressing the paradox pointed out by Einstein, Podolsky and Rosen  (EPR) \cite{EPR}. The EPR paradox arose because of the  counter-intuitive
non-locality of quantum mechanics. It was meant as thought experiment to explicitly to demonstrate the incompleteness of the nonclassical theory. In order to address the controversy that ensued between EPR, on one side, and Bohr on the other side, Bell derived a set of inequalities \cite{Bell} which should be obeyed if reality was  local and entanglement did not exist. The violation of Bell inequalities by quantum mechanics could thus demonstrated experimentally, which eventually was accomplished in a series of pioneering experiments carried out by the team led by Aspect~\cite{aspect_epr}. Up to this point in time,  the majority of experiments show the correctness of quantum mechanics and therefore the existence of entanglement.  More recently,  entanglement has been realized to be an important resource for quantum computation and quantum communication~\cite{NielsonChuang}. The exploitation of entangled pair as an ebit (entangled  qubit)  can speed up quantum computation and communication. Indeed, some quantum information processes such as quantum teleportation rely heavily on the use of ebits. 

In the last two decades,  quantum information-theoretic concepts have also had a strong influence on many fundamental  aspects of  condensed matter theory, statistical mechanics, and quantum field theory \cite{Amico}. This so-called many-body theory has used entanglement as a new tool to study  condensed matter theory, especially for  the numerical calculations of strongly correlated system at zero temperature.  The success of density matrix renormalization group
(DMRG) \cite{DMRGBook}  and other methods based on the tensor network such as matrix product states (MPS)  in  one dimension \cite{MPSR,DMRGBook} lies on the fact that the
entanglement between subsystems in one dimension is essentially small \cite{EntCrit}.  The understanding of the entanglement has   provided new insights, for example, into critical phenomena, where it has been shown that entanglement can diverge just like the susceptibility at a
second-order critical point \cite{CriticalEnt},   and the scaling of the entanglement entropy (see below for a definition)  can provide a new way to calculate the central
charge~\cite{EntCrit}.  

 As mentioned above, entanglement can also be applied to gain deeper understanding of quench dynamics, and in particular, the emergence of the GGE. 
The point  of view of how the GGE  emerges from the correlations between the eigenmodes of the (post-quench) Hamiltonian was arrived at beginning with the work on the LM~\cite{cazalilla2006}, and culminating with the work reported in Ref.~\cite{CazalillaIucciChung}, which involved the authors of the present work.  
In order to introduce the main ideas of Ref.~\cite{CazalillaIucciChung} and their application to
quenches in the LM, let us start by reviewing some results about reduce density matrices. For a bipartite system where the system is divided into a subsystem part, $A$, and an environment part, $B$, a density matrix $\rho_0= |\Psi_{AB}\rangle \langle \Psi_{AB} | $ can be obtained from a \emph{pure} state  $|\Psi_{AB}\rangle$ describing the composite $AB$ system. The reduced
density matrix $\rho_A$ obtained by tracing out  the environment:
\begin{equation}
    \rho_A = \tr_B \rho_0 = \tr_B |\Psi_{AB}\rangle \langle \Psi_{AB} |.
\end{equation}
The Hermitian operator $\rho_A$ is interesting for various reasons.
First, it allows to obtain the von Neumann entanglement entropy of the subsystem $A$ by means of the expression:
\begin{equation} \label{Sent}
   S_A  = -\tr \rho_A \log{\rho_A}.
\end{equation}
This measure of entanglement is central to quantum information theory. One of the reasons is that,  as shown by Wooters and coworkers, under local operations and classical communication (LOCC, i.e. a local unitary transformation),  an entangled state of a bi-partite system can only be transformed into a state with the same or lower entanglement entropy~\cite{Wooters}. In addition, reduced density matrices are used to efficiently truncate the Hilbert space basis for the density matrix renormalization group based methods \cite{DMRGBook,tddmrg}.  An important result about the entanglement entropy is the way  
 it scales with the size of the subsystem $A$.  The area law \cite{AreaLaw} states that for a subsystem of dimension $d$, $S_A(d) \propto L^{d-1}$ 
 (for critical systems a logarithmic correction also appears).  The area law is the fundamental reason why DMRG is so efficient in 1D since the entanglement entropy grows at most as $\log{L}$. 
 
 Rather than focusing on the entanglement entropy,  much more structure can be found in the eigenvalues of the reduced density matrix, namely, the entanglement spectrum. The latter can be obtained by diagonalizing the entanglement
Hamiltonian, $H_{A}^{ent}$ \cite{EntSpect,RDMPeschel}, which is defined through the  relation: 
\begin{equation} \label{eq:entH}
   \rho_A = \frac{e^{-H_A^{\mathrm{ent}}}}{\tr{e^{-H_A^{\mathrm{ent}}}}}.
\end{equation}
Note that the basis that diagonalizes $H_A^{\mathrm{ent}}$ also diagonalizes
$\rho_A$. Next,  let us  focus our discussion of entanglement spectra 
of systems described by a  Hamiltonian 
that is quadratic in terms of certain (fermionic or bosonic) quasi-particle 
operators (as is the case of the LM), $b_{\gamma}$ and $b^{\dag}_{\gamma}$, i.e.
\begin{equation}
   H_0 = \sum_{\gamma \delta} \left[ A_{\gamma \delta} b_{\gamma}^{\dagger}
   b_{\delta} + \frac{1}{2} (B_{\gamma \delta} b_{\gamma}^{\dagger}
 b_{\delta}^{\dagger} + \mathrm{h.c.}) \right]
\end{equation}
In the above expression $\gamma, \delta$ are the quasi-particle
quantum numbers, which can be coordinates,  wave vectors, etc.
The reduced density of such models can be obtained by first
obtaining a coherent state representation of the the matrix elements of the full density matrix $\rho_0$~\cite{ChungPeschel} and then explicitly 
integrating out the degrees of freedom of the environment. In this way
the reduced density matrix $\rho_A$ can be separated as a direct
product form \cite{RDMPeschel, GFM}:
\begin{equation} \label{rhoAform}
  \rho_A = \frac{e^{- \sum_m \lambda_m {\alpha}_m^{\dagger} {\alpha}_m}}{\tr e^{- \sum_m \lambda_m {\alpha}_m^{\dagger} {\alpha}_m}}
 \end{equation}  
where ${\alpha}_m^{\dagger }$ (${\alpha}_m$)  are the creation
(annihilation) operators that diagonalize the entanglement Hamiltonian, $H^{ent}_{A}$.  The single-particle entanglement spectrum is given by
\begin{equation}
     \lambda_m = \log{\frac{1 \pm \mu_m}{\mu_m}},\label{eq:ents}
\end{equation}
where  $\mu_m$ are the eigenvalues of the block correlation function matrix
\begin{equation}
      G_{\gamma \delta} = \tr \rho_0 \: b_{\gamma}^{\dagger} b_{\delta},
\end{equation}
where the quantum numbers $\alpha$ and $\beta$ are  restricted to the subsystem $A$. The plus (minus) sign in Eq.~\eqref{eq:ents} applies  to bosons (fermions). We will see below, in
section~\ref{subsec:LMEP}, that 
the GGE can be obtained as product of reduced density matrices of the form 
of Eq.~\eqref{rhoAform}.  

%Finally, it is worth mentioning that,  in terms of the single-particle entanglement spectrum, the entanglement
%entropy of the subsystem can be written as:
%
%\begin{equation}
%S_A = -\sum_{m} \left[ \lambda_m \log \lambda_m \pm (1 \mp  \lambda_m)  \log %\left(1 \mp \lambda_m \right) \right]. 
%\end{equation}
%  

 The above results apply only to the case that the state of the system  can be described by a Gaussian density matrix, $\rho_0$, which in the pure state case can be regarded as the ground state of a quadratic Hamiltonian. Otherwise the block correlation function matrix is not enough to describe the entanglement  due to the failure of the Wick theorem.  The results for the entanglement spectrum also apply  to mixed Gaussian density matrices, such as, for example a thermal density matrix $\rho_0 \propto e^{-H_0/T}$, where 
$T$ is the absolute temperature and  $H_0$ is a quadratic Hamiltonian. However,
in this case, the expression for von Neumann entropy of $\rho_A$ 
Eq.~\eqref{Sent} cannot be used  to calculate the entanglement due to   the additional thermal contribution to the entropy. 

\subsection{GGE and the steady state of the Luttinger Model} 
\label{subsec:LMEP}

In this subsection, we shall consider a quantum quench in the Luttinger model
(LM) and show why the GGE provides a description of the asymptotic steady state at $t\to +\infty$ from the
perspective of entanglement~\cite{CazalillaIucciChung}. The initial state of the quench is
assumed to a gaussian state $\rho_0 \propto e^{-H_0/T}$ (the pure state case is obtained by letting $T\to 0$). A general form for the pre-quench Hamiltonian $H_0$ is:
\begin{eqnarray}
  H_0 & =  & \sum_{q, q^{\prime}} [\epsilon_0(q) \delta_{q, q^{\prime}} +
  V_0(q, q^{\prime})] b^{\dagger}(q) b(q^{\prime}) \nonumber \\
  &  &+   \sum_{q, q^{\prime}}  [\Delta_0^{\star}({q, q^{\prime}})
  b(q) b(q^{\prime}) + \Delta_0({q, q^{\prime}})  b^{\dagger}(q^{\prime}) b^{\dagger}(q) ].
\end{eqnarray}
To keep our discussion general and connect to the discussion in the previous section, we shall
consider that $H_0$ is not translational invariant and couples different wave numbers by means of the potentials $V_0(q,q^{\prime})$ and $\Delta_0(q,q^{\prime})$. Therefore the initial state $\rho_0$ breaks the translational invariance of the system. We assume a sudden quench where at $t=0$, the Hamiltonian becomes diagonal in the modes described by  $b(q)$ and $b^{\dag}(q)$ (cf. $H_{LM}$ in Eq.~\ref{eq:hambos}). 

Gaussian initial states like $\rho_0$ have the important property that   Wick's theorem  allows us to obtain the correlators of an arbitrary product of   $b(q)$ and $b^{\dagger}(q)$ eigenmode operators  from two-point correlation functions, e.g. $\langle b^{\dagger}(q) b(q^{\prime}) \rangle_0 = \tr \rho_0
b^{\dagger}(q) b(q^{\prime}),  \langle b(q) b(q^{\prime}) \rangle_0$ and
$\langle b^{\dagger}(q) b^{\dagger}(q^{\prime}) \rangle$, et cetera.  In addition, another useful property of Gaussian states, which was described in the previous section, is that the reduced density matrices of an arbitrary partition of the system are also Gaussian. In particular, if we choose a partition where the sub-system $A$ is one of the modes, say, $q$ and the environment $B$ is the rest $q^{\prime}\neq q$, then, tracing the environment yields
\begin{equation}
   \rho(q) = \tr_B \rho_0 =  \tr_{q\neq q^{\prime}} \rho_0 =  \frac{e^{-\lambda(q) I(q)}}{Z(q)},
\end{equation} 
where $I(q) = b^{\dagger}(q) b(q)$ is the quasiparticle
occupation operator. For this particular partition the entanglement
Hamiltonian equals $\lambda(q) I(q)$ and 
$\lambda(q)$ is the single-mode entanglement spectrum, which 
related to the occupation number  of the density matrix (cf. Eq.~\ref{eq:ents}), 
$ n(q) = \langle I(q) \rangle = \tr \rho_0 I(q) = \tr \rho_0  b^{\dagger}(q)b(q) $ by means of the relation:
\begin{equation}
\lambda(q) = \log{\frac{1 \pm n(q)}{n(q)}}. 
\end{equation}
The claim (to be substantiated below) is that the GGE can be constructed as a product of such single-model reduced density matrices, i.e.
\begin{equation}
    \rho_{\mathrm{GGE}} = \bigotimes_q \rho(q).  
\end{equation}
Indeed, for the relevant local and non-local operators, decoherence erases the dependence of the correlators  on the off-diagonal eigenmode correlations~\cite{CazalillaIucciChung} of the type $\langle b^{\dag}(q) b(q^{\prime}) \rangle$ (for $q\neq q^{\prime}$),  $\langle b(q) b(q^{\prime}) \rangle$, etc.  
Thus, if all relevant correlators depend only on the $n(q) = \langle I(q)\rangle$, taking the expectation value over the GGE and over $\rho_0$ yield the same
result. However, this point of view of the GGE is tantamount to the mathematical statement that each eigenmode acquires a mode-dependent effective temperature $T(q) = \lambda(q)/\left[ v(q)|q|\right]$ as a result of its entanglement with
other eigenmodes in the initial state of the system. 

Next, we discuss how the GGE emerges in the LM.   In order to make connection with the previous discussion about quenches of the interaction,  we shall consider in the following translational invariant initial states. Thus, we focus on gaussian states that are obtained from the initial Hamiltonians of the form:
\begin{equation} \label{eqn:H0}
     H_0 = \sum_{q \neq 0}  \left\{  v_0(q)  |q|  b^{\dagger}(q) b(q) + \frac{1}{2} g_0(q) |q|  \left[b^{\dagger}(q) b^{\dagger}(-q) +b(q) b(-q)\right] \right\}, 
\end{equation} 
which respects translational invariance. For a quench of the interaction, $v_0(q) =v_F \cosh{2\varphi(q)} $ and $g_0(q) = v_F \sinh{2\varphi(q)}$, as follows from Eq.~\eqref{eq:bogoleq} ($\varphi(q)$ is the Bogoliubov angle). Furthermore, since the observables in which we are interested 
can be expressed in terms of the boson field  $\phi_{\alpha = R,L}(x)$ (cf. Eq.~\eqref{eq:phib}),  we rewrite this operator in terms of the eigenmodes of post-quench Hamiltonian,  $H_{LM}$, which yields:
\begin{align} 
     \Phi_{\alpha}(x,t) &=   \sum_{q>0} \left( \frac{2 \pi}{q
   L}\right)^{\frac{1}{2}}  e^{i s_{\alpha} q} [\cosh{\varphi(q)}
      e^{-i v(q) |q|t}        
      b(s_{\alpha} q) - \sinh{\varphi(q)}  e^{i v(q) |q|t}
      b^{\dagger} (-s_{\alpha} q)], \\
      \phi_{\alpha}(x) &= s_{\alpha}\phi_{\alpha} + \frac{2\pi x}{L}  N_{\alpha}       + \Phi_{\alpha}(x) + \Phi^{\dag}_{\alpha}(x). \label{eq:phie}
\end{align}
In the LM and for gaussian states like $\rho_0$, correlation functions of vertex operators can be expressed in term of two-point correlations of the fields $\phi_{\alpha}(x,t)$. For the vertex operators, the key identity allowing for the evaluation of the vertex-operator correlation functions is the following:
\begin{equation}   \langle e^{i A_{\alpha}(x_1,\dots,x_n,t )} \rangle_0 = \tr \left[ \rho_0 e^{i A_{\alpha}(x_1,\dots,x_n,t )}  \right] = e^{-\frac{1}{2}
 \langle A_{\alpha}^2(x_1,\dots,x_n,t) \rangle_0},  \label{eq:Aalpha}
\end{equation}
where 
\begin{equation} 
     A_{\alpha}(x_1,\dots,x_n,t) = \sum_{i=1}^n p_i
     \phi_{\alpha}(x_i,t), 
 \end{equation}
with $\sum p_i =0$. The proof of this identity relies on Wick's theorem~\footnote{Eq.~\eqref{eq:Aalpha} can be proven by
expanding in series the exponential in the left-hand side in a Taylor
series and applying Wick's theorem to all the terms, which involve powers of
$A(x_1, \dots ,x_n)$. Resuming the resulting series, the right  hand-side of Eq.~\eqref{eq:Aalpha} is obtained}.  By the same token,  correlation functions of $\rho_{\alpha}(x) = \partial_x \phi_{\alpha}(x)/2\pi$ can be obtained. Note that $\phi_{\alpha}(x)$ itself is not an observable. However, observables in the LM are related to correlation functions of  $\rho_{\alpha}(x)$ and vertex operators. Despite this fact, the correlations of $\phi_{\alpha}(x)$ play a central role in the LM, as we have just shown.  For example, using Eq.~\eqref{eq:Aalpha}, the two-point correlation function of the right moving Fermi fields
\begin{align}
C_{\psi_{R}}(x,t)  &=   \langle \psi_R^{\dagger}(x,t) \psi_R(0,t) \rangle_0 =
\tr \left[\rho_0 \psi_R^{\dagger}(x,t) \psi_R(0,t) \right]\\
 &=  \mathcal{A} \exp{[C_{\phi_R}(x,t)  - C_{\phi_R}(0,t) ]},\label{eq:vr}
\end{align} 
where $\mathcal{A}$ is a cut-off dependent prefactor. Thus, because of Wick's theorem, it is sufficient to consider the two-point correlations of  $\phi_{\alpha}(x,t)$, which, using \eqref{eq:phie}, can be
written as follows:  
\begin{equation}
    C_{\phi_R}(x,t) = \langle \phi_R(x,t) \phi_R(0,t) \rangle_0 =
    D_{\phi_R}(x) +   F_{\phi_R}(x,t), 
\end{equation} 
where
\begin{equation} 
D_{\phi_R}(x) = \sum_{q \neq 0}  \left|\frac{\pi}{q L} \right|
[\cosh{2\varphi(q)}  + \operatorname{sgn} (q)]  \left\{e^{i q x} [1 + n(q)]
+ e^{-i q x} n(q)\right\}
\end{equation}
is the contribution of the diagonal correlations of the eigenmodes 
in the initial state $n(q) = \langle
b^{\dagger}(q) b(q)\rangle_0$. Furthermore,
\begin{equation}    
 F_{\phi_R}(x,t)  = \sum_{q \neq 0}  \left|\frac{\pi}{q L} \right|
[\cosh{2\varphi(q)}  + \operatorname{sgn} (q)] \left[e^{iqx -2iv(q)|q|t} \Delta(q) + e^{−iq x+2iv(q)|q|t}
\Delta^{\star}(q)\right], 
 \end{equation}
where $\Delta(q) = \langle b(q) b(-q) \rangle_0$ is related to the anomalous
correlations of the eigenmodes in the initial state. 
Due to the translation invariance,  $n(q)$ and $\Delta(q)$  are the only 
non-vanishing two-point correlations of the eigenmodes in the initial state. 

At this point, we are ready to show that the  correlation
functions of the LM  only depend on the diagonal correlations, 
$n(q)$.  To this end, we notice that, whereas the contribution
of the diagonal correlations $n(q)$ is time independent, the
contribution of the anomalous terms depends on time. Hence, 
because of dephasing between the different
Fourier components (mathematically, by the Riemann-Lebesgue lemma),
in the thermodynamic limit, the contribution of $F_{\phi_R}(x,t)$ vanishes  as 
$t \rightarrow +\infty$. Explicitly, at zero
temperature ($T=0$), for the quench of the interaction in the LM, 
\begin{equation}
    F_{\phi_R}(x,t) - F_{\phi_R}(0,t) \sim \log{\left|
        \frac{(2vt)^2 - x^2}{(2vt)^2}\right| },
\end{equation}
which vanishes as $t\to +\infty$ (provided $x$ is kept finite). This implies that all correlations are
asymptotically determined by $D_{\phi_R}(x)$, which depends only on
$n(q) = \langle b^{\dagger}(q) b(q) \rangle_0$. This observation
allows to trace out all the $q^{\prime}\neq q$ eigenmodes since 
$n(q) =\tr \left[ \rho_0 b^{\dag}(q) b(q) \right] 
= \tr [\rho(q) b^{\dagger}(q) b(q)] = \tr
\rho_{\mathrm{GGE}}  b^{\dagger}(q) b(q)$. Thus, we arrive at the same result as if we had used the GGE density matrix $\rho_{\mathrm{GGE}} = \bigotimes_q \rho(q)$.  Hence,
$C_{\phi_R}(x,t\to +\infty) - C_{\phi_R}(x,0) = 
C^{GGE}_{\phi_R}(x)  - C^{GGE}_{\phi_R}(0)$,
where $C^{GGE}_{\phi_R} = \tr \left[\rho_{\mathrm{GGE}} \phi_R(x) \phi_R(0) \right]$,
and using Eq.~\eqref{eq:vr} yields:
\begin{equation}
\lim_{t\to +\infty} C_{\psi_R}(x,t) = C^{GGE}_{\psi_R}(x).
\end{equation}

 It is worth noting that the translational invariant initial state implies 
that eigenmode correlations are bipartite, that is, each mode at $q$ is
entangled only with the eigenmode at $-q$ (see next section). Thus, we can regard the effective temperature $T(q)$ for the eigenmodes with $q$ as the result of their quantum correlations with the $-q$ eigenmodes and vice versa.  However, the translational invariance of initial states is not a necessary condition for the long time correlations to be described by the GGE. If the translational-invariance constraint is relaxed, 
dephasing  will still erase the off diagonal correlations, not only the ``anomalous'' ones $\langle b(q) b(q^{\prime}) \rangle_0$ but also the normal ones $\langle b^{\dag}(q) b(q^{\prime})\rangle$ ($q^{\prime}\neq q$) because they always appear in the correlators multiplied by phase factors  of the form $e^{i\left[v(q)|q|\pm v(q^{\prime})|q^{\prime}|\right]| t}$, which oscillate very rapidly for $t \to \infty$ and therefore yield a vanishing contribution.  
This is essentially the reason why the GGE is so effective for describing the asymptotic state correlations. Furthermore, it also shows why only the occupation operators of the eigenmodes (quasi-particles) of the post-quench Hamiltonian, i.e., $I(q) = b^{\dag}(q) b(q)$, are the only integrals of motion required for its construction. 

Similar ideas has been extended in Ref.~\cite{CazalillaIucciChung} to other exactly solvable models in one dimension, such as the quantum Ising model and the XX model in order to show that, in the thermodynamic limit, they relax to the GGE. They have been also applied to understand the the pre-thermalized state
of a 2D Fermi gas in terms of the GGE by Nessi, Iucci, and Cazalilla~\cite{nessi2014} (cf.  section~\ref{sec:pretherm}).

\subsection{Entanglement spectra from generalized Gibbs}

In the previous section we have discussed how the GGE emerges from dephasing and the diagonal correlations between the eigenmodes. We have pointed that LM  is a
system with bipartite eigenmode entanglement due to the coupling of the right moving ($q>0$) and the left moving ($q<0$) Tomonaga bosons that is mediated through the interaction. For systems with bipartite entanglement, relaxation to the GGE can be discussed on general grounds. 
This interpretation  might allow for the  experimental possibility to measure the entanglement spectrum by studying the steady state of post-quench correlations\cite{chung2012}. 

 Let us consider a general system consisting of  two subsystems $A$
and $B$. For $t > 0$, the Hamiltonian of the system is quenched to a Hamiltonian of the form:
 \begin{equation} \label{eq:HAB}
  H_0=H_A + H_B + H_{AB}
 \end{equation}
where $H_A, H_B$ and $H_{AB}$
are quadratic in some eigenmodes ${\alpha}(q), \beta(q)$ which carry a
quantum number $q$ and can be bosonic or fermionic (at this point we consider both, for the sake of generality), i.e.
\begin{eqnarray}
    H_A & = & \sum_q \varepsilon_A(q) \alpha^{\dagger}(q) \alpha(q), \\
    H_B & = & \sum_q \varepsilon_B(q) \beta^{\dagger}(q) \beta(q) \\
    H_{AB} & = & \sum_q \Delta_{AB}(q)  \left[ \alpha^{\dagger}(q)
      \beta(q) + \beta^{\dagger}(q) \alpha(q)  \right]. 
\end{eqnarray} 
We assume that the system is prepared in a thermal  initial density
matrix ($\rho_0 = Z_0^{-1} e^{-H_0/T}$). For $t>0$, the
coupling between the two systems $H_{AB}$ disappears, and the two
subsystems evolve unitarily  and are uncoupled according the
Hamiltonian $H = H_A+H_B$. The existence of the coupling $H_{AB} $ for
all $t \leq 0$ means that in the initial state $\rho_0$, there are
correlations (i.e. bipartite entanglement) between the eigen modes,
i.e. $\langle \alpha^{\dagger}(q) \beta(q) \rangle \neq 0$. 

From the discussion in the last subsection, we have seen that the GGE is equivalent to an effective description of correlations that, in the asymptotic long time steady state, only depend on the diagonal correlations of the eigenmodes. The latter are entirely parametrized by an effective temperature,
resulting from entanglement with the other modes, and which determines the reduced density matrix of the eigenmode. When the correlations are bi-partite, we can regard the effective temperature for the modes in the subsystem
$A$  as due to the entanglement with the modes in the subsystem $B$
(and vice versa). Thus whenever we are dealing with $\langle
\alpha^{\dagger}(q) \alpha(q) \rangle = \tr \rho_0 \alpha^{\dagger}(q) \alpha(q) $
and $\langle
\beta^{\dagger}(q) \beta(q) \rangle = \tr \rho_0 \beta^{\dagger}(q) \beta(q) $
we can trace out one of the subsystems and obtain
\begin{eqnarray}
    n^{\alpha}(q) &=& \langle \alpha^{\dagger}(q) \alpha(q) \rangle
    =\tr \rho_A \alpha^{\dagger}(q) \alpha(q) =
    \rho_{\mathrm{GGE}} \alpha^{\dagger}(q) \alpha(q), \\ 
  n^{\beta}(q) &=& \langle \beta^{\dagger}(q) \beta(q) \rangle
    =\tr \rho_A \beta^{\dagger}(q) \beta(q) =
    \rho_{\mathrm{GGE}} \beta^{\dagger}(q) \beta(q), 
\end{eqnarray}
where $\rho_A = \tr_B \rho_0$ and $\rho_B = \tr_A \rho_0$. Therefore
$\rho_{\mathrm{GGE}}$ is written as 
\begin{equation} \label{eq:GGEDM}
   \rho_{\mathrm{GGE}} = Z^{-1}_{\mathrm{GGE}}
   \exp{\left\{ -\sum_q \left[\lambda_{\alpha}(q)  \alpha^{\dagger}(q) \alpha(a) +
   \lambda_{\beta}(q)   \beta^{\dagger}(q) \beta(q)\right] \right\}}
\end{equation}
with the Langrange multipliers $\lambda_{\alpha}(q) =\log{[{1\pm
  n^{\alpha}(q)}/n^{\alpha}(q)]}$ and $\lambda_{\beta}(q) =\log{[{1\pm
  n^{\beta}(q)}/n^{\beta}(q)]}$ (plus sign for bosons and minus sign
for fermions).  Therefore the GGE density matrix can be written as 
\begin{equation}
   \rho_{\mathrm{GGE}} = \rho_A \otimes \rho_B. 
\end{equation}
We regard the result as a way to related the density matrix of GGE to
the reduced density matrix, and hence to the entanglement
Hamiltonian by Eq.~\eqref{eq:entH}. Thus we see $\rho_{\mathrm{GGE}}$
is determined by the total entanglement Hamiltonian as 
$H = H_A^{ent} + H_B^{ent}$, where 
\begin{eqnarray}
    H_A^{ent} & = &\sum_q \lambda_{\alpha}(q)   \alpha^{\dagger}(q)
    \alpha(q) =\sum_q \log{\left[(1\pm
        n^{\alpha}(q))/n^{\alpha}(q)\right]} \alpha^{\dagger}(q) \alpha(q) \\
    H_B^{ent} & = & \sum_q \lambda_{\beta}(q)   \beta^{\dagger}(q)
    \beta(q)  = \sum_q \log{\left[(1\pm
        n^{\beta}(q))/n^{\beta}(q)\right]} \beta^{\dagger}(q) \beta(q), 
\end{eqnarray}
which, by comparison with Eq.~\eqref{eq:GGEDM}, allows us to identify
$\lambda_{\alpha}(q)$ and $\lambda_{\beta}(q)$ as the entanglement
spectrum of the subsystems $A$ and $B$. 

Therefore the entanglement spectra determine the asymptotic state
following a quantum quench. We can therefore obtain the entanglement
spectra and entropy by measuring the behavior after a quench
process. In the following we show that the entanglement spectra of the
LM can be accessed by the quantum sudden quench. 

For the sake of definiteness, let us next consider  initial states corresponding to  
constant $v_0$ and $g_0(q) = (\Delta/2\pi)$. Therefore,  the initial
Hamiltonian reads
\begin{equation} \label{eq:HLM}
   H_0 = \sum_{q>0} \left[  v_0 q (\alpha^{\dagger}(q)
     \alpha(q) + \beta^{\dagger}(q) \beta(q) ) + \frac{\Delta}{2\pi}
     q (\alpha^{\dagger}(q)
     \beta ^{\dagger}(q) + \beta(q) \alpha(q) )\right]
\end{equation} 
with the redefinition of the operators $\alpha(q) = b(q)$  and
$\beta(q) = b(-q)$. Essentially Eq.(\ref{eq:HLM}) and
Eq.(\ref{eq:HAB}) are the same by applying particle hole
transformation for the system $B$. 

 Eq.\eqref{eq:HLM} is diagonalized by a canonical transformation where $a(q) = \cosh{\varphi (q)} \alpha(q) -\sinh{\varphi(q)} \beta^{\dagger}(q)$, $a(-q) = -\sinh{\varphi(q)} \alpha(q) +\cosh{\varphi(q)}
\beta^{\dagger}(q)$ and choosing $\tanh{2\varphi(q)} = -\Delta/(2\pi v_0)$, the
initial Hamiltonian reads:
\begin{equation}
    H_0 = \sum_{q>0}  v(q) q (a^{\dagger}(q) a(q)+
    a^{\dagger}(-q)  a(-q) ) 
\end{equation} 
where $v(q) =  v_0 \sqrt{(1-(\Delta/2\pi v_0)^2)}$. 
In order to obtain the entanglement spectra $\lambda_{\alpha}$ and
$\lambda_{\beta}$, the occupation number are used:
$n^{\alpha}(q) = n^{\beta}(q) = \sinh^2{\varphi(q)} =
1/2( v_0 /v(q) -1)$. Hence 
\begin{equation}
   \lambda_{\alpha} = \lambda_{\beta} =  \log{\left( \frac{v_0  +
         v(q)}{v_0-v(q)} \right)} = \varepsilon,
\end{equation}
where $\varepsilon = 2[\log(2\pi  v_0)/\Delta + \sqrt{(2\pi
   v_0/\Delta)^2-1}]$ is a constant. Therefore the entanglement
Hamiltonian $H_A^{ent} = \sum_q \varepsilon  \alpha^{\dagger}(q)
\alpha(q)$ and $H_B^{ent} = \sum_q \varepsilon \beta^{\dagger}(q)
\beta(q)$. 
The lowest entanglement eigenvalue is $0$. There is also  a `flat band' of  entaglement eigenvalues of $\varepsilon$. This kind of spectra may be extracted by the experimental determination of the eigenmode dependent temperatures that
parameterize the GGE in an interaction quench of the LM.

\subsection{Non-gaussian initial states}

From the discussions above the Gaussian initial state $\rho_0$ is needed to prove GGE correct. On the contrary, Dinh,　 Bagrets, and Mirlin (DBM) \cite{Mirlin} studied a sudden quench of the interaction in LM assuming  a double-step initial momentum distribution function of the fermions (a situation relevant to experiments in the quantum Hall regime of the two-dimensional electron gas, see section~\ref{sec:expmeso} and references). Using non-equilibrium bosonization, they obtained the steady state energy ($\propto$ momentum) distribution. DBM pointed out that the resulting steady state distribution cannot
be obtained from the GGE. The latter, at large distances,  predicts an exponential decay of single-particle density matrix, i.e.~\cite{Mirlin}:
\begin{equation}
   C_{\psi R}(x) \approx C_{\psi R}^{(0)}(x) \left| \frac{R}{x}
   \right|^{-\alpha} e^{-\kappa |x|}, 
\end{equation}
where $\alpha$ and $\kappa$ depend on the interaction and details of the initial state. The reason why GGE fails in this case is because the initial state contains a non-Gaussian correlations amongst the  bosonic eigemodes of the system. This kind of non-Gaussian memory survives in the steady state. A similar conclusion has been reached by Sotiriadis for general initial non-Gaussian states  using conformal field theory methods~\cite{Sotiriadis}.

\section{Brief survey of Luttinger's relatives}\label{sec:other}

Various kinds of perturbations to the LM  have been
considered as well as their various effects on the quench
dynamics. The number of possible perturbations is rather large,
and given the space constraints, we cannot make justice to all
the recent developments in this area. We merely mention the 
most relevant here. 

\subsection{Quenches in the sine-Gordon model}

 A well known perturbation to the LM is the sine-Gordon model , e.g. 
\begin{equation}
H_{SGM} = H_{TLL} + g(t) \int dx \cos 2 \phi(x), \label{eq:sg}
\end{equation}
where $H_{TLL}$ is the fixed-point Hamiltonian 
in Eq.~\eqref{eq:tllham}. Assuming $g(t) = g \theta(t)$, for instance, this model describes  the sudden application
to the LM of an \emph{external} periodic potential that is
commensurate with half the Fermi wave number $\pi/p_F$ (recall $p_F$ is the Fermi momentum)~\cite{cazalilla_rmp,giamarchi_book}.  The model has also a dual version where $2\phi \to \theta$ in the cosine term, which describes  a quench of the Josephson coupling~\cite{giamarchi_book,cazalilla_rmp}.  In equilibrium (i.e. for a time-independent coupling $g(t)$) the cosine perturbation
is relevant in the renormalization group sense 
for $K \lesssim 2$ (for infinitesimal $g$)~\cite{bosonization_new,giamarchi_book,cazalilla_rmp},
which opens a spectral gap. For $K > 2$, the perturbation 
is irrelevant and the low energy spectrum is thus gapless and adiabatically connected to the LM spectrum (up to corrections that rapidly decrease with the excitation energy). 

Like the LM, quantum quenches in the sine-Gordon model have
also attracted much attention. Iucci and Cazalilla~\cite{iucci2010} studied a sudden quench of the cosine term 
where $g(t) = g \theta(-t)$ in the so-called harmonic limit (holding for $K \ll 1$) and at the so-called Luther-Emery line~\cite{lutheremery1974} (corresponding to $K = 1$ \cite{cazalilla_rmp} for Eq.~\ref{eq:sg}). The results are entirely consistent  with the general results of Cardy and Calabrese~\cite{cardycalabrese2006} for a quench from an off critical to a critical Hamiltonian. Iucci and Cazalilla also 
showed that in both the harmonic limit and the Luther-Emery line the system relaxes to the GGE.
In addition, the reverse quench (from critical 
to non-critical) was also analyzed in Ref.~\cite{iucci2010}.  Applications of the quench of the sine-Gordon to the experiments in Schmiedmayer's group have been discussed also recently by Dalla Torre, Demler, and Polkovnikov~\cite{dallatorre}, and by Foini and Giamarchi~\cite{foini}.

 Smooth quantum quenches where the coupling $g(t) \sim t^{r}$ have been studied by De Grandi, Gritsev, and Polkovnikov (GGP)~\cite{degrandi}, who focused on the dynamics near (i.e. starting from or ending at) the critical point between the gapped and gapless phases. By changing the exponent $r$, it is possible to interpolate between  the sudden quench ($r \to 0$) and the adiabatic quench
limit (for $r\to \infty$). In between, for the linear quench $r=1$, the Kibble-Zurek mechanism can be studied.  Rather than the dynamics of correlations, GGP focused on the production rate of excitations, $P_{ex}$, the  density of the quasiparticles $n_{ex}$, the diagonal entropy $S_d$ and the heat (the excess energy above the new ground state of the post-quench Hamiltonian) $Q$. They showed that the scaling of $P_{ex}$, $n_{ex}$ and $S_d$ are associated with the singularities of the generalized adiabatic susceptibility $\chi_{2r+2}(\lambda)$ of order $2r+2$ defined as
\begin{equation}
   \chi_m(\lambda)  = \frac{1}{L^d} \sum_{n\neq 0} \frac{|\langle n| V | 0\rangle|^2}{[E_n(\lambda)-E_0(\lambda)]^m} 
\end{equation}  
where a $d$-dimensional perturbative Hamiltonian $H(\lambda) = H_0 + \lambda V$ is considered with the eigenenergy $E_n$ of the state $|n\rangle$, while if the quench ends at the critical point the scaling of $Q$ is related to $\chi_{2r+1}$~\cite{degrandi}. For a sudden quench, i.e. $r=0$, $\chi_{2r+2}(\lambda)$ is reduced to the fidelity susceptibility $\chi_{2}  \equiv \chi_f$. In  two exactly solvable limits: the massive bosons (i.e. the harmonic limit) and the massive fermions, they  also obtained  results for quenches at  finite temperature. Due to the statistics of the quasiparticles, they showed that the structure of the singularity remains the same except that for $n_{ex}$ and $Q$,  the dimensionality $d$ is replaced by $d-z$, where $z$ is  dynamical exponent for bosons. On the other hand,  for fermions $d\rightarrow d+z$. The difference stems from the bunching of bosons, which enhances non-adiabatic effects, whereas anti-bunching of fermions suppresses transitions~\cite{degrandi}. 
\subsection{Long-ranged hoping models}

Other systems that are attracting much interest in recent times in connection with experiments in ion traps are models with long-ranged interactions (see e.g. Ref.~\cite{iontraps}). We have already discussed how long-ranged interactions affect the post-quench correlations of the LM~\cite{nessi2013}. Other types of interactions may correspond to a long-range hoping of bosons in a lattice, which translates into a long-ranged Heisenberg exchange for spins. Tezuka, Garc\'{\i}a-Garc\'{\i}a and Cazalilla (TGC) studied a quench of the range of the boson hoping, focusing on the dynamics of the condensate~\cite{tezuka2014}. When hoping amplitude decays as a power-law of the distance $r$, i.e. $t_{r}\sim |r|^{-\kappa}$,
the system  exhibits long range order at \emph{zero temperature} 
for $\kappa < 3$ (up to interaction-induced corrections)~\cite{cazalilladis,lobos}. Quenching the power-law tail of the hoping amplitude (or, equivalently, the value of $\kappa$) is tantamount to changing the effective dimensionality  of the system~\cite{tezuka2014}. Using
bosonization~\cite{cazalilla_rmp}, TGC obtained that the condensate fraction (normalized to the initial state fraction) $f(t)$ decays at short times as $f(t) = 1 - b t^2$
and at long times as a stretched exponential $f(t)\sim 
e^{- c t^{ \frac{(3-\kappa)}{2}}}$,
where $b$ and $c$ depend on the model parameters like lattice filling, interaction, etc. These predictions were found to be in reasonable  agreement~\cite{tezuka2014} with numerical results obtained using 
td-DMRG, despite the fact that the bosonization treatment does
not take into account the possibility of phase slips, which may
be required in order to achieve a complete understanding of the
dynamical destruction of the condensate following the quench of the hoping range.

\subsection{To thermalize or to not thermalize}

There is a great deal of evidence  that observables of
generic, non-integrable isolated systems relax to a state that can described by a standard thermal equilibrium ensemble~\cite{rigol_eth1,rigol_eth2} (see also Ref.~\cite{review_rigol} and references therein). Thus,
for such a generic systems,  the LM results that we have reviewed above should break down at sufficient long times.
Correlations are therefore expected to crossover to their thermal averages (e.g. Eq.~\ref{eq:corrthermal}). Thermalization being impossible to avoid in most cases, the question  is therefore one of relaxation dynamics and time scales. And the latter can be very long
due to the limited available phase space for quasi-particle scattering in especially one-dimensional systems~\footnote{Of course, this assumes the validity of some kind of quasi-particle picture, as it is the case of the Luttinger model, but may not be case of other critical models.}. 

 In the intermediate time-regime, before the crossover to thermal behavior takes place, the LM predictions should be accurate provided the system is not strongly excited at the outset in the quench process. This expectation is based on the observation that, even if the high excited states of a generic, non-integrable model are highly chaotic~\cite{review_rigol}, the low-energy part of spectrum may still retain some features that can be captured by a suitable exactly solvable model like the LM. 

It becomes therefore apparent that the results described above  should describe some kind of pre-thermal state, which exists for some time, and which crosses over at a later time to a fully thermal state. How this happens and what kind of perturbations to the LM drive such thermalization is still very much under debate (although a number of important results have emerged recently, see below). 

A first attempt to understand how terms that have been neglected by the use of TLL fixed-point (i.e. the LM) Hamiltonian to describe the quench dynamics was undertaken by Mitra and Giamarchi~\cite{mitra}. Using the non-equilibrium Green's function (Keldysh) formalism, and assuming that a non-linearity in the form of a sine-Gordon term (cf. Eq.~\ref{eq:sg}) is adiabatically switched-on following an interaction quench, they showed that the system would eventually reach a thermal state. In other words, the coupling between the eigenmodes due to the sine-Gordon term introduces quasi-particle scattering that violates the infinite conservation laws of the LM and relaxes the system to a thermal state. 

 Nevertheless, intrinsic sources of scattering between quasi-particles are present in most models of the TLL class at all times. One of them is the curvature in the fermion dispersion, which　in bosonized form reads~\cite{haldanejpc}:
\begin{equation}
H_{m} = \frac{1}{m} \int dx \left(\partial \theta \right)^2 
\partial_x \phi.\label{eq:curv}
\end{equation}
The effects on the post-quench dynamics of the \emph{resonant} scattering of Tomonaga bosons caused 
by Eq.~\ref{eq:curv} have been recently addressed by Buchhold, Heyl, and Diehl (BHD)~\cite{diehl}. By using non-equilibrium  Green's functions,  BHD wrote a quantum kinetic equation for the Tomonaga bosons in the presence of collisions mediated by the Eq.~\eqref{eq:curv}. From the numerical solution of the kinetic equation, the following picture for the fermion single-particle correlation function emerges:
Besides the two regimes that have been discussed in
section~\ref{sec:dyn},
which are called pre-quench and pre-thermal by the authors of  Ref.~\cite{diehl}, a new regime, termed `thermal'
appears.  In the thermal regime,  $C_{\psi_R}(x,t)$ exhibits an exponential decay with distance. Thus,
summarizing BHD's results for a quench of the interaction starting from the non-interacting ground state, the following three distinct regimes exist:
\begin{equation}
C_{\psi_R}(x,t) = \left\{
\begin{array}{cc}
 Z(t) C^{(0)}_{\psi_R}(x,t), \quad  Z(t) \sim t^{-\gamma^2} & 
0 < t < t_x = |x|/2v, \\
C^{\mathrm{GGE}}_{\psi_R}(x) \simeq \left|\frac{R}{x} \right|^{\gamma^2}  &  t_x < t < t_{\mathrm{th}}\\
e^{-|x|/\xi_{\mathrm{th}}[T(t)]} & t > t_{\mathrm{th}}.
\end{array}
\right. 
\end{equation}
In the  pre-quench regime (i.e. for $t < t_x$), correlations are only multiplicatively modified from their initial state values.
The pre-thermalized regime corresponds to  $t > t_x = 2 v t$. In this  regime, the dynamics is controlled by  the exactly solvable
truncation of the total Hamiltonian, i.e. $H_{LM}$. 
In this regard,  $H_{LM}$ plays a similar role
 to the exactly solvable truncation of the
interacting 2D Fermi gas discussed in section~\ref{sec:pretherm}. In other words, the exactly solvable models describe a regime of `inertial response' to the quench of the interaction, in which quasi-particles are formed before they can start to scatter each other.  

The final (``thermal'') regime is takes places for $t > t_{\mathrm{th}}$. The thermalization time  $t_{\mathrm{th}} \simeq   t_0 (x/R)^{1/\alpha(K)}$,  where $0 < \alpha(K) < 1$  and $\beta(K)$ are functions of the Luttinger parameter, $K$, that need
to be determined numerically~\cite{diehl} ($R$ is the interaction range, which effectively plays the role of momentum cut-off, see section~\ref{sec:history}). The curvature, Eq.~\eqref{eq:curv}, leads to the emergence of a new time scale in the problem $t_0 = R^2/w_0 \sim m R^2$, where  $w_0 \sim 1/m\sqrt{K}$ is the strength of the interaction-vertex arising from Eq.~\eqref{eq:curv}. BHD also pointed that the existence of the thermalized
regime and the different scaling of the time (or equivalently, length) scales that determine them, 
implies the existence of a minimum time or distance below which the pre-thermalized behavior cannot be observed. 
This distance, $x_{\mathrm{m}}$ is found by equating $2 v t_{\mathrm{m}}=\beta(K) R (t/t_0)^{\alpha(K)}$~\cite{diehl} and hence $x_{\mathrm{m}} = 2 v t_{\mathrm{m}}$, and $|x| < x_m$ the quasi-particles with short wavelength have not formed before they begin to scatter each other. Thus, below this
length scale the system correlations will behave either
as in the pre-quench or in the thermal regime. 

 Finally, it is worth pointing out that the thermal regime is a stationary state only asymptotically. This is because the correlation length 
 $\xi_{\mathrm{th}}[T(t)] = K/(1+K^2) \left(v/\pi T(t)\right) $, where
 the effective temperature $T(t) = T+ \Delta(K) (v/R) (t/t_0)^{-\mu}$, where $T$ is the final temperature of the system,  $\Delta(K)$ is a parameter that is determined numerically, and the exponent $mu = 2/3$~\cite{diehl}.

\section{Relevance to experiments}\label{sec:exp}

\subsection{Ultracold atomic gases}

In Ref.~\cite{cazalilla2006}, the experimental realization that was envisaged was a quantum quench in a  dipolar Fermi gas effectively confined to one dimension in a very anisotropic trap~\footnote{Note that, unlike for sthe Coulomb potential, the Fourier transform of the dipolar potential is regular in one dimension.}. However, cooling Fermi gases to  temperatures well below the Fermi temperature, $T_F$, is technically
difficult, although some progress has been recently reported in the case of dipolar Fermi gases due to their long-ranged interactions~\cite{dipolar_fermigases}.  Currently, the lowest temperatures attainable are $\approx 10\%$ of $T_F$. Thus, even if the dynamics of trapped ultracold gases can be  studied for quite some time due to their isolated nature making it extremely quantum coherent, finite temperature effects must be accounted for when comparing with the experiment. Such effects were theoretically addressed in 
Ref.~\cite{iucci2009} for the momentum distribution, with the conclusion that the latter is probably not the most ideal observable to study quench dynamics. As argued in Ref.~\cite{iucci2009}, the reason are the rather small differences between the non-equilibrium steady state momentum distribution and the 
interacting finite temperature momentum distribution may be hard to discern experimentally. Let us recall that the, at finite temperatures, the discontinuity in momentum distribution is absent due to entropic effects. In addition, the discontinuity is also very sensitive to other effects, such as
inhomogeneity, finite-size, etc. 
However, it is worth noticing that not all finite-temperature effects are perverse. Indeed,  if the initial state is a finite temperate state $\rho_0 = e^{-H_0/T}$, then the steady state is reached  much faster,  $t \approx 1/T$~\cite{cazalilla2006,iucci2009},
rather than for $t \to +\infty$. 

Cooling problems are less severe for ultracold atomic gases of bosons, and therefore much more progress has been made in studying the non-equilibrium dynamics of such systems. In fact,  using a trapped 1D cloud of bosons that suddenly is split longitudinally into two 1D clouds, it has been possible to 
observe  pre-thermalization~\cite{pretherm1,pretherm2} and  also find strong evidence for relaxation to the generalized Gibbs ensemble (in the pre-thermalized regime)~\cite{schmiedmayer}. As we have mentioned in the previous section, to some extent,  these
experiments can be interpreted  in terms of a quench in a sine-Gordon model that is dual  to the one  in Eq.~\eqref{eq:sg}, i.e. with  
the replacement $\cos 2\phi \to \cos \theta$~\cite{iucci2010,dallatorre,foini}. In the experiment, the field $\theta$ corresponds to relative phase between the two atomic clouds and the coupling $g(t)$, which describes the Josephson tunneling between the two  clouds, is suddenly quenched from a large to a zero value. In other words, the quench proceeds from the gapped to the gapless phase of the sine-Gordon model. Since the initial (Josephson) coupling $g(t < 0)$ is large due to the initial complete overlap of the two clouds, and the atoms in the 1D clouds are weakly interacting, the  harmonic approximation where $\cos \theta \approx -\theta^2/2$ is a good starting point~\cite{pretherm1,pretherm2,iucci2010,dallatorre,foini}. Thus, the quench can be described by a quadratic model and a gaussian initial state, and it is therefore expected to relax to the GGE, as we have discussed in section~\ref{sec:entgge}. 
In the experiment, the post-quench correlations of the (relative) phase  are measured by letting the two atomic clouds interfere after releasing them from the trap~\cite{schmiedmayer,pretherm1,pretherm2}.  
\begin{figure}
\centering
\includegraphics[scale=0.5]{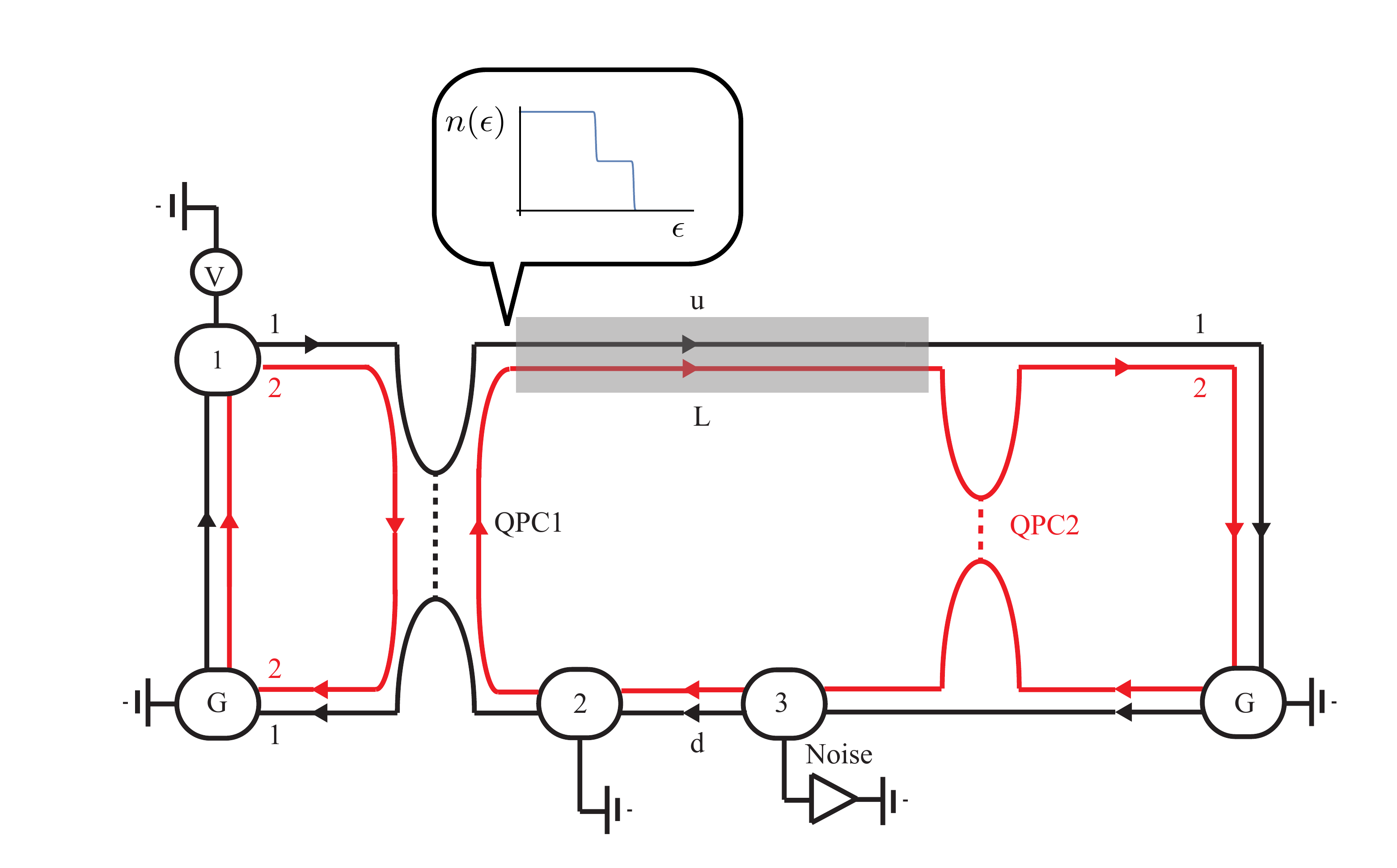}
\caption{Schematic diagram of the device used to prepare an non-equilibrium state and probe its relaxation in the $\nu  = 2$ quantum Hall state of a 2D
electron gas subject to a magnetic field.}
\label{fig:device}
\end{figure}

\subsection{Mesoscopic systems}\label{sec:expmeso}

Despite being ultracold atomic gases a major motivation for the study of quantum quenches and non-equilibrium dynamics, it may appear as somewhat striking to some that,  some 
of the above theoretical ideas have found faster applications in the realm of mesoscopic systems of electrons. However, when subjected to intense magnetic fields and low temperatures, the ``dirty'' 2D gases of interacting electrons in the quantum Hall regime have been for some time the arena for many experiments that have reached the precision standards of atomic physics in interferometry\cite{yang2002}, for instance.

 Chirality arising from the application of the magnetic field makes the edges of the 2D electron gas in the \emph{integer} quantum Hall  regime behave as clean, non-interacting, conducting channels as far as the transport at small voltage bias is concerned. This is manifested in the perfect conductance quantization of the latter in units of $e^2/h$, which can be experimentally observed at low temperatures. However, this does not 
mean that the (screened) Coulomb interaction can be neglected. Coulomb interaction is indeed important to account for the spectral properties of the 
channels. Furthermore, its effect is especially felt when driving the system out of  equilibrium. In this regard,   Kovrizhin and Chalker~\cite{chalker1,chalker2} 
first pointed out the interesting analogies between the study of the relaxation mechanisms of integer quantum Hall interferometers and the interaction quench in the LM. In this section, we shall closely follow the work of Milletar\`i and Rosenow~\cite{milletari}, which has also found recent experimental confirmation~\cite{inue}.

  Using a Hall bar device like the one shown in 
 Fig.~\ref{fig:device} at Landau-level filling $\nu = 2$, it is possible to prepare a double-step non-equilibrium  distribution for the fermions in the outer edge channel (denoted by $1$ in Fig.~\ref{fig:device}). Mathematically,
 \begin{equation}
 n_{1R}(\epsilon) = a \theta(\mu_1 -\epsilon) + (1-a) \theta(\mu_2-\epsilon),\label{eq:doublestep}
 \end{equation}
where $\mu_1 = (1-a) eV$ and $\mu_2 = -a eV$ ($eV > 0$),
and $0 < a < 1$.
The initial state corresponding to the above distribution  is prepared by taking the outer channel $1$, which comes from a reservoir at chemical potential $eV$ through the first quantum point contact (QPC1). At this contact, the channel is brought in close contact with its grounded left-moving partner at the other edge of the Hall bar, which allows for local tunneling of electrons with different chemical potentials between the two.  For non-interacting electrons, this results in the distribution function displayed in Eq.~\eqref{eq:doublestep}. After being driven out of equilibrium, channel $1$ is allowed to relax by interacting downstream with the grounded inner channel $2$ in the shaded region of Fig.~\ref{fig:device}. Chirality implies that time and space play the same role and the spatial overlap of the two channels in this region can be regarded as sudden quench of their mutual interaction. In the interaction region the channels are
described by the (`post-quench')  Hamiltonian:
\begin{equation}
H =  \sum_{q > 0} \left( 
a^{\dag}_{1R}(q)\, a^{\dag}_{2R}(q)\right)
\left( 
\begin{array}{cc}
v_1 q & g_{4} q/2\\
g_{4} q/2 & v_{2} q 
\end{array}
\right)  \left(
\begin{array}{c}
a_{1R}(q)\\
a_{2R}(q)
\end{array}
\right), 
\end{equation}
which can be diagonalized
by means of the canonical transformation  
$a_{1R}(q) =  \cos\theta  \, b_{1R}(q) - \sin \theta \, b_{2R}(q)$ and
$a_{2R}(q) = \sin \theta \, b_{1R}(q) + \cos \theta \, b_{2R}(q)$. The 
 $\theta$ is the mixing angle, which is
determined from $\tan 2\theta  = \frac{g_4}{v_1-v_2}$, where $g_4$ parametrizes the (screened Coulomb) interaction between the two channels and  $v_1$ and $v_2$ are the outer and inner channel velocities, respectively. Note that the appearance of trigonometric functions here (instead of the hyperbolic functions of the previous sections) is due to the same (right-moving) chirality of the two coupled fermionic channels. Nevertheless,  as in the case of the LM, a solution of the quench dynamics can be obtained by means of the following time-dependent canonical transformation:
\begin{align}
a_{1R}(q,t) &= f_1(q,t) a_{1R}(q) + g(q,t) a_{2R}(q),\\
a_{2R}(q,t) &= g(q,t) a_{1R}(q) + f_{2}(q,t) a_{2R}(q),
\end{align}
where 
\begin{align}
f_{1}(q,t) &= \frac{1}{2}\left( e^{-i \tilde{v}_1 q t} + e^{-i\tilde{v}_2 q t} \right)
+  \frac{1}{2} \left( e^{-i \tilde{v}_1 q t} - e^{-i \tilde{v}_2 q t} \right) \cos 
2 \theta, \\
f_{2}(q,t) &= \frac{1}{2}\left( e^{i \tilde{v}_1 q t} + e^{i \tilde{v}_2 q t} \right)
-  \frac{1}{2} \left( e^{i \tilde{v}_1 q t} - e^{-i \tilde{v}_2 q t} \right) \cos  2 \theta, \\ 
g(q,t) &=  \frac{1}{2} \left( e^{-i \tilde{v}_1 q t} - e^{-i \tilde{v}_2 q t} \right) \sin  2 \theta.
\end{align}
The parameters  $\tilde{v}_{1(2)} = v_{1(2)}\cos^2 \theta + v_{2(1)} \sin^2 \theta \pm \frac{1}{2} g_{4} \sin 2\theta$ are
the eigenmode velocities.   
\begin{figure}
\centering
\includegraphics[scale=1.0]{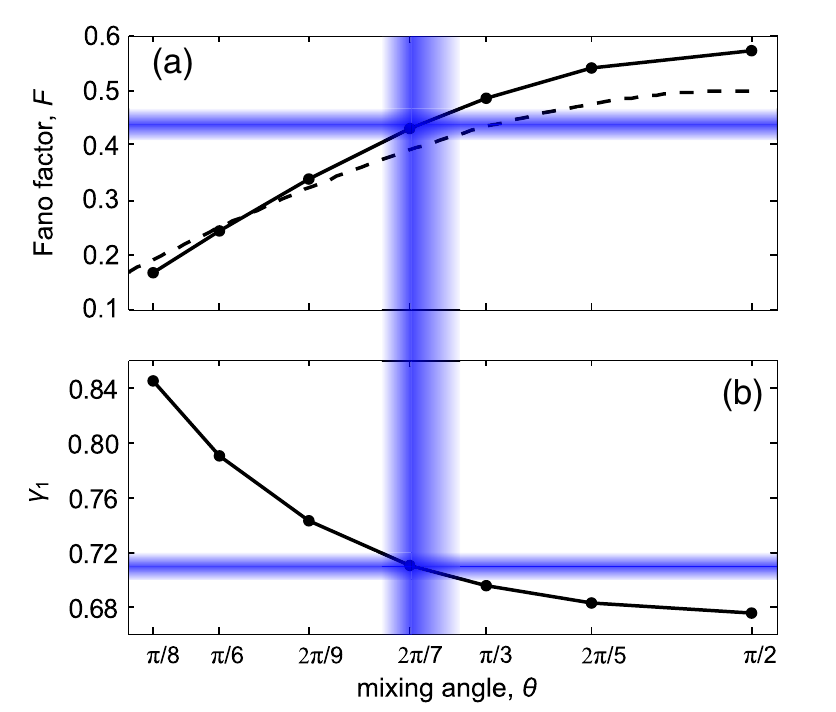}
\caption{(a) Fano factor representing the fractional charge $e^{\star} = F e$ in the inner channel $2$ (cf. Fig.~\ref{fig:device})  vs. the mixing angle $\theta$, which measures the strength of the inter-channel interaction. 
(b) Exponent $\gamma_1$ of the excess noise in the inner channel $2$ which is fitted acording to $\left[a (1 -a) \right]^{\gamma_1}$, where $a$ is the transmission probability through the first quantum point contact (QPC1, see Fig.~\ref{fig:device}). The dots in both figures correspond to the numerical results obtained from the theory of Ref.~\cite{milletari}. The dashed line is a simple model proposed in Ref.~\cite{inue}, for which $F = (\sin \theta)/2$. The shaded (blue) area represents the best fit with the error of $1.2\sigma$ based on the fitted exponent $\gamma_1$.}
\label{fig:inue_exp}
\end{figure}

After spatially overlapping, the two channels $1$ and $2$ are again spatially separated in order to probe the inner channel $2$  by taking it through  a second quantum point contact at QPC2. There, it comes into contact with its left-moving partner before reaching ground together with outer channel (cf. Fig.~
\ref{fig:device}). The shot noise in this
channel is thus measured at contact number $3$~
(cf. \ref{fig:device}). Roughly speaking, shot
noise measures the (fractional) charge of the carriers
and can be obtained from the single-particle Green's 
function of the channel ($\alpha=R,L$):
\begin{equation}
G^{<}_{2\alpha}(\tau) = \langle \psi^{\dag}_{2\alpha}(x,t + \tau)
\psi_{2\alpha}(x,t)\rangle.
\end{equation}
through the expression:
\begin{equation}
S = \frac{2e^2}{\hbar} |t_2|^2\int d\epsilon 
\left[ G^{<}_{2R}(\epsilon) G^{<}_{2L}(-\epsilon) + 
 G^{<}_{2L}(\epsilon) G^{<}_{2R}(-\epsilon)  \right] 
\end{equation}
where $t_2$ is the tunneling amplitude at the second point contact,
QPC2. The Green's function of the inner upstream channel can be written as follows:
\begin{align}
G^{<}_{2R}(\tau) &= G^{(0,<)}_{2R}(0,\tau) Z(x, t,\tau),\\
Z(x, t,\tau) &=\langle e^{i\chi^{\dag}(x,t,\tau)} e^{-i\chi(x,t,\tau)} \rangle, \\
G^{(0,<)}_{2R}(0,\tau) &= 
\frac{1}{2\pi} \frac{1}{\left(-i\tilde{v}_{1}\tau + a_0\right)^{\sin^2 \theta} \left( -i \tilde{v}_2 \tau + a_0 \right)^{\cos^2 \theta}}, 
\end{align}
where $\chi(x,t,\tau) = \sum_{q > 0} \left(\frac{2\pi}{q L} \right)^{1/2} e^{-q a_0/2} \: e^{i q x_0} \left[g(q,t+\tau) - g(q,t) \right] a_{1R}(q)$ and $a_0$ is a short-distance  cut-off of the order of the magnetic length.
The calculation of $Z(x,t,\tau)$ is complicated by non-gaussianity of the initial state, which yields the double-step distribution of Eq.~\eqref{eq:doublestep}.
However, it can be still expressed in terms of a Fredholm determinant of the Toeplitz type and evaluated numerically~\cite{milletari}. The properties of the steady state are accessed by taking the limit $t \to +\infty$ in the above Green's function. The resulting steady state is not thermal~\cite{milletari}, which is a consequence of the constrained dynamics of the model. The shot noise can be then extracted from the steady state Green's function. The so-called Fano factor, corresponding to the ratio $F = S/S_{\mathrm{ref}}$ where 
$S_{\mathrm{ref}} = 4 e p a (1 -a)(e^2/h) V$ ($p =  |t_2|^2 \tilde{v}^{\sin^2\theta}_1 \tilde{v}^{\cos^2 \theta}_{2} $ is the reflection probability at the second quantum point contact~\cite{milletari}, and $V$ is the voltage bias),  is shown in  Fig.~\ref{fig:inue_exp} as a function of the mixing angle, $\theta$. The experimental
determination of its value (with the corresponding error bars)  is indicated  by the shaded (blue) region. 

\section{Conclusions and outlook}\label{sec:summ}

Having the opportunity to write this article has taught 
us that the subject of quantum quenches in the Luttinger and related models continues to enjoy tremendous vitality
after ten years. There are plenty of new analytical results, often obtained by the combination of powerful techniques which include conformal field theory, Bethe-ansatz, non-equilibrium (Keldysh) Green's functions, etc. 
In addition, powerful numerical methods have been applied to various models and are teaching us where the
analytics can be (and cannot be) applied.

 Nevertheless, a deeper understanding of the universality of the LM predictions, beyond the rather fragmented knowledge obtained from specific models, is still lacking. In the authors' view, such a framework still needs to be put in place. This framework should relate to some of the concepts of entanglement that have been surveyed in section~\ref{sec:entgge} and should be able to account for the differences between gaussian and non-gaussian initial states. More importantly, the theoretical framework 
should also provide \emph{quantitative} answers
to questions about time scales and requirements
for thermalization. It should tell which perturbations
to the Hamiltonian of an exactly solvable model like Luttinger's are dominant in driving the system towards thermal equilibrium,  which ones subdominant, and what the time scales associated with them are. 

On the experimental side, results are coming 
out at an increased pace. These include the recent observation of relaxation to the generalized Gibbs ensemble in an ultracold gas~\cite{schmiedmayer}. However, as 
theorists who cherish their own (theoretical) pets, 
we would like to see similar experiments carried out
for a more faithful (fermionic) realization of the interaction quench in the Luttinger model. Our hope is that
this article may motivate a young experimentalist to
take on this challenge. In addition, let us mention that using different probes in the mesoscopic
setup discussed in section~\ref{sec:expmeso} may allow
for a more thorough characterization of the non-equilibrium steady state, and whether it is really a steady state or just a pre-thermalized one. Ultracold atomic systems still have a lot of offer as well as trapped ions~\cite{iontraps} since possibilities for the control of the system parameters and the preparation of the initial state 
are larger for these systems. 

There are also new venues waiting to be fully explored. There is a zoo of newly discovered topologically-protected phases out there. Many of them are endowed with gapless states at their interfaces with topologically trivial matter. In two dimensions, some of these gapless states can be described as Tomonaga-Luttinger liquids, and the ideas described in section~\ref{sec:expmeso} can be relevant for the understanding of their non-equilibrium dynamics as well. From a broader perspective, we may also wonder what kind of new manifestations in the non-equilibrium dynamics of interacting system topological protection can bring in.  

Going beyond isolated systems (or systems that can be treated to a large extent as such), the dynamics of the
Luttiger model and its relatives coupled to environments is, to the best of our  knowledge, a largely uncharted territory. The study of mesoscopic systems is clearly an area that can benefit from such studies since, in the solid state, it is much more difficult to completely isolate fragile quantum systems from their environment. Accounting for the effects of realistic sources of quantum dissipation~\cite{cazalilladis}  is therefore 
an interesting research direction. Furthermore, in ultracold atomic systems,  despite being largely isolated,  it is also possible to engineer interesting environments~\cite{eneko} and the study of quantum dynamics in those setups should be a promising research topic.

\acknowledgments 

We gratefully acknowledge A. Iucci, A. M. Garc\'{\i}a-
Garc\'{\i}a, N. Nessi, and  M. Tezuka for collaborations and discussions. Throughout the years, we have also benefited from discussions  with many colleagues, particularly with L. Amico, E. Demler, R. Fazio, T. Giamarchi, S. Kehrein, M. Milletar\`{\i}, A. Mirlin, A. Muramatsu, G. Mussardo, I. Peschel, A. Polkovnikov, M. Rigol,  L. Santos, M. Haque, and D. Weiss.   MAC also thanks M. Milletar\`{\i} for his help in producing Fig.~\ref{fig:device} and useful comments on section~\ref{sec:expmeso}. Funding from the Ministry of Science and Technology (MoST, Taiwan) and the National Center for Theoretical Sciences (MAC) is also gratefully acknowledged. This work is dedicated to the memory of Alejandro Muramatsu
without whose inspiring conversations this work would not have happened.


\begin{thebibliography}{30}
%
\bibitem{editorial}
M. Rigol and M. A. Cazalilla, New J. of Phys. 
{\bf 12},  055006 (2012).
%
\bibitem{review_polkovnikov}
A. Polkovnikov, K. Sengupta, A. Silva, and M. Vengalattore,
Rev. Mod. Phys. {\bf 83} 863 (2011). 
%
\bibitem{KZMReview} J. Dziarmaga, Adv. in Phys. {\bf 59}, 1063 (2010)
%
\bibitem{review_rigol}
L. D'Alessio, Y. Kafri, A. Polkovnikov, and M. Rigol
arxiv:1509.06411 (2015).
%
\bibitem{luttinger1963}
J.~M. Luttinger, J. Math. Phys. {\bf 4}, 1154 (1963).
%
\bibitem{mattislieb1965}
D.~C. Mattis and E.~H. Lieb, J. Math. Phys. {\bf 6}, 304 (1965).
%
\bibitem{jordan_neutrino}
P. Jordan, Z. Phys. {\bf 93} 464 (1935); \emph{ibid} {\bf 98}, 795 (1936);
\emph{ibid} {\bf 99}, 109 (1936);  \emph{ibid} {\bf 105}, 114 (1937);
\emph{ibid} {\bf 105} 229 (1937).
%
\bibitem{tomonaga1950}
S. Tomonaga,  Progr.~Theor.~Phys.~(Kyoto) {\bf 5}, 544 (1950).
%
\bibitem{schwinger1959}
J. Schwinger, Phys.~Rev.~Lett. {\bf 3}, 296 (1959).
%
\bibitem{little1964}
W.~A. Little, Phys. Rev. {\bf 134}, A1416 (1964).
%
\bibitem{lutheremery1974}
A. Luther and V.~J. Luther, Phys.~Rev.~Lett. {\bf 33}, 589 (1974).
%
\bibitem{lutherpeschel1975}
A. Luther and I. Peschel, Phys.~Rev.~B {\bf 12} 3908, (1975).
%
\bibitem{haldane_tll}
F.~D.~M. Haldane, Phys.~Rev.~Lett.~ {\bf 45}, 1358 (1980).
%
\bibitem{haldanejpc}
F.~D.~M. Haldane, J. of Phys. C {\bf 14}, 2585 (1981).
%
\bibitem{haldane1981}
F.~D.~M. Haldane, Phys.~Rev.~Lett. {\bf 47}, 1840 (1981) 
%
\bibitem{bosonization_old}
K.~D. Schotte and U. Schotte, Phys. Rev. {\bf 182}, 479 (1969);
D. C. Mattis, J. Math. Phys. {\bf 15}, 609 (1974);
S. Mandelstam, Phys. Rev. D {\bf 11}, 3026 (1975).
%
\bibitem{bosonization_new}
A. A. Gogolin, A. A. Nersesyan, and A. M. Tsvelik, \emph{Bosonization of Strongly Correlated Electrons}, Cambridge (2002).
%
\bibitem{giamarchi_book}
 T. Giamarchi \emph{Quantum Physics in One Dimension} (Oxford 2006).
%
\bibitem{rigoletal2006}
M. Rigol, V. Dunjo, V. Yurovskii, and M. Olshanii, Phys.~Rev.~Lett. (2006).
%
\bibitem{kinoshita2006}
T. Kinoshita, T. Wenger, and D. S. Weiss,  Nature (London)
{\bf 440}, 900 (2006).
%
\bibitem{cazalilla2006}
M.~A. Cazalilla, Phys.~Rev.~Lett.~{\bf 97} 156403 (2006). 
%
\bibitem{mitra}
A. Mitra and T.Giamarchi, Phys. Rev. Lett. {\bf 107}, 150602 (2011)
%
\bibitem{liebrobinson}
E. H. Lieb and D. Robinson, Commun. Math. Phys. {\bf 28}, 251, (1972).
%
\bibitem{cardycalabrese2006}
J. Cardy and P. Calabrese, Phys.~Rev.~Lett. {\bf 96}, 136801 (2006).
%
\bibitem{iucci2009}
A. Iucci and M.~A. Cazalilla, Phys.~Rev.~A {\bf 80}, 063619 (2009).
%
\bibitem{nessi2013}
N. Nessi and A. Iucci, Physical Review B {\bf 87}, 085137 (2013). 
%
\bibitem{tddmrg}
M. A. Cazalilla and J.~B. Marston,  Phys.~Rev.~Lett. {\bf 88}, 256403 (2002);
G. Vidal, Phys. Rev. Lett. {\bf 91} 147902 (2003);
S.~R. White and A. E. Feiguin, Phys. Rev. Lett. {\bf 93}, 076401 (2004);
U. Sch\"ollwock, Rev. Mod. Phys. {\bf 77}, 259 (2005);
U. Sch\"ollwock, and S.~R. White, {\it Methods for Time Dependence in DMRG}
in G. G. Batrouni and D. Poilblanc (eds.) {\it Effective models for low-dimensional strongly correlated systems}, p. 155, AIP, Melville, 
New York (2006).
%
\bibitem{cazalilla2004}
M.~A. Cazalilla, J. Phys. B {\bf 37} S1 (2004).
%
\bibitem{rg_shankar}
R. Shankar, Rev. Mod. Phys. {\bf 66}, 129 (1994). 
%
\bibitem{rg_book}
See e.g. S. Sachdev, \emph{Quantum Phase Transitions}, 2nd edition, 
Cambridge University Press (Cambridge, UK, 2011) and also Refs.~\cite{bosonization_new,giamarchi_book}.
%
\bibitem{cazalilla_rmp}
M. A. Cazalilla, R. Citro,  T. Giamarchi, E. Orignac, and M. Rigol, 
Rev.~Mod.~Phys.~{\bf 83}, 1405 (2010). 
%
\bibitem{karraschmedem2012}
C. Karrasch, J. Rentrop, D. Schuricht, V. Meden, Phys.~Rev.~Lett.~{\bf 109}, 126406 (2012).
%
\bibitem{collura2015}
M. Collura, P. Calabrese, and F.~H.~L. Essler, Phys.~Rev.~ B  {\bf 92}, 125131 (2015).
%
\bibitem{berges}
J.Berges, S. Bors\'anyi, and C. Wetterich,
Phys.Rev. Lett. {\bf 93}, 142002 (2004).
%
\bibitem{pretherm1}
T. Kitagawa, A. Imambekov, J. Schmiedmayer, and
E. Demler, New J. Phys. {\bf 13}, 073018 (2011).
\bibitem{pretherm2}
M. Gring, M. Kuhnert, T. Langen, T. Kitagawa, B. Rauer, M. Schreitl, I. Mazets, D. A. Smith, E. Demler, and J. Schmiedmayer, Science {\bf 337}, 1318 (2012).
%
\bibitem{moeckel2008}
M. Moeckel and S. Kehrein, Phys. Rev. Lett. 100, 175702 (2008); Ann. Phys. (N. Y.) {\bf 324}, 2146 (2009).
%
\bibitem{eckstein2009}
M. Eckstein, M. Kollar, and P. Werner, Phys. Rev. Lett. {\bf 103}, 056403 (2009).
%
\bibitem{nessi2014}
N. Nessi, A. Iucci, and M. A. Cazalilla, 
Phys. Rev. Lett. {\bf 113}, 210402 (2014).
%
\bibitem{FSbosonization}
A. Houghton and J. B. Marston, Phys. Rev. B {\bf 48}, 7790
(1993); A. Houghton, H.-J. Kwon, and J. B. Marston, Phys.
Rev. B {\bf 50}, 1351 (1994); A. Houghton, H.-J. Kwon, and J. B. Marston, Adv. Phys. {\bf 149}, 141 (2000).
%
\bibitem{chung2012}
M.-C. Chung, A. Iucci, and M. A. Cazalilla, New J. Phys. {\bf 14}, 075013 (2012).
%
\bibitem{EPR} 
A. Einstein, B. Podolsky and N. Rosen, Phys. Rev. 
{\bf 47}, 777 (1935). 
% % % 
\bibitem{Bell} 
J.~S. Bell  Physics {\bf 1} (3): 195–200  (1964).
% % %
\bibitem{NielsonChuang} 
M. A. Nielsen and I. L. Chuang 
{\it Quantum Computation and Quantum Information}, 
Cambridge University Press (Cambridge, UK, 2000). 
%  
\bibitem{Amico} 
L. Amico, R. Fazio, A. Osterloh and V. Vedral,
  Rev. Mod. Phys. {\bf 80}, 517 (2008).
%
\bibitem{aspect_epr}
A. Aspect, Ph. Grangier, and G. Roger, 
Phys.~Rev.~Lett. {\bf 49}, 91  (1982). 
%
\bibitem{DMRGBook} I. Peschel, X. Wang, M  Kaulke and K.  Hallberg,
  {\it Density-Matrix Renormalization}, Springer 
% % %
\bibitem{MPSR} F. Verstraete, J. I. Cirac and V. Murg, Adv. Phys. {\bf
    57},143 (2008).
% % %
\bibitem{EntCrit} G. Vidal, J. I. Latorre, E. Rico and A. Kitaev, Phys.Rev.Lett. {\bf 90}, 227902,(2003)
% % %
\bibitem{CriticalEnt} A. Osterloh, L. Amico, G. Falci and R. Fazio,
  Nature, {\bf 416}, 609 (2002)
% % % 
\bibitem{EntSpect} H. Li and F. D. M. Haldane, Phys. Rev. Lett. {\bf
    101}, 010504 (2008).
% 
% % %
\bibitem{Rigol} M. Rigol, V. Dunjko, V.  Yurovsky and M. Olshanii,  Phys. Rev. Lett. {\bf 98} 050405; 
M.  Rigol M, A. Muramatsu and M. Olshanii, Phys. Rev. A {\bf 74},  053616 (2006)
% % %
\bibitem{CazalillaIucciChung} M. A. Cazalilla, A. Iucci and
    M.-C. Chung, Phys. Rev.  E {\bf 85}, 011133 (2012)
% % %
\bibitem{Wooters} C. H. Bennett, D. P. DiVincenzo, J. A. Smolin and
    W. K. Wooters, Phys. Rev. A {\bf 54}, 3824 (1996)
% % %
\bibitem{AreaLaw}  J. Eisert, M. Cramer and M.B. Plenio,
    Rev. Mod. Phys. {\bf 82} (1). 277 (2010).
%   
\bibitem{jaynes}
E.~T. Jaynes, Phys. Rev. {\bf 106}, 620 (1957);\emph{ibid} {\bf 108}, 171 (1957).  
% % %
\bibitem{RDMPeschel} I. Peschel and  V. Eisler, J. Phys. A:
  Math. Theor.{\bf 42} 504003 (2009)
% % %
\bibitem{rigolgge2}
M. Rigol, A. Muramatsu, and M. Olshanii, Phys. Rev. A {\bf 74}, 053616 (2006).
%
\bibitem{rigolgge3}
A.~C. Cassidy, C.~W. Clark, Marcos Rigol,
Phys. Rev. Lett. {\bf 106}, 140405 (2011).
%
\bibitem{ChungPeschel} 
M. C. Chung and I. Peschel, Phys. Rev. B {\bf 64}, 064412 (2001). 
% % %
\bibitem{GFM} 
S.-A. Cheong and C. L. Henley, Phys. Rev. B {\bf 69}, 075111 (2004); I. Peschel, J. Phys. A {\bf 36}, L205 (2003); T. Barthel, M.-C. Chung, and U. Schollw\"ock, Phys. Rev. A {\bf 74}, 022329 (2006).
% % %
\bibitem{DoraHaqueZarand} 
B. Dora, M. Haque and G. Zar{\'a}nd,
    Phys. Rev. Lett. {\bf 106}, 156406 (2011)
% % %
\bibitem{LE} B. Dora, F. Pollmann, J. Fort{\'a}gh and G. Zar{\'a}nd,
    Phys. Rev. Lett. {\bf 111}, 046402 (2013)
% % %
\bibitem{anderson_oc}
P. W. Anderson, Phys. Rev. Lett. {\bf 18}  1049 (1967).
%
\bibitem{fidelity_qpt}
F. Pollmann, S. Mukerjee, A. G. Green, and J. E. Moore,
Phys. Rev. E {\bf 81}, 020101 (2010).
%
\bibitem{fidelity}
J. O. Fjaerestad, J. Stat. Mech.  P07011 (2008);
M.-F. Yang, Phys. Rev. B {\bf 76}, 180403 (2007).
% % %
\bibitem{Orignac} J.-S. Bernier, R. Citro, C. Kollath and E. Orignac,  Phys. Rev. Lett. {\bf 112}, 065301 (2014).
% % % 
\bibitem{Dziarmaga} J. Dziarmaga and M. Tylutki, Phys. Rev. E {\bf  84}, 214522 (2011) 
% % %
\bibitem{Mirlin} S. N. Dinh, D. A. Bagrets and A. D. Mirlin,
  Phys. Rev. B {\bf 88}, 245405 (2013)
% % %
\bibitem{Sotiriadis} S. Sotiriadis, arXiv: 1507.07915 (2015).
%
\bibitem{dipolar_fermigases}
K. Aikawa, A. Firsch, M. Mark, S. Baier, R. Grimm, and 
F. Ferlaino, Phys.~Rev.~Lett. {\bf 112}, 010404 (2014); 
N.~Q. Burdick, K. Baumann, Y. Tang, M. Lu, B. L. Lev, Phys.~Rev.~Lett. {\bf 114}, 023201 (20125.
%
%
\bibitem{schmiedmayer}
T. Langen \emph{et al.}  Science {\bf 348}  207 (2015).
%
\bibitem{iucci2010}
A. Iucci and M. A. Cazalilla, 
New J. Phys. {\bf 12} 055019 (2010).
%
\bibitem{dallatorre}
E. Dalla Torre, E. Demler, and A. Polkovnikov, Phys. Rev. Lett. 110, 090404 (2013)
%
\bibitem{foini}
L. Foini and T. Giamarchi, Phys. Rev. A {\bf 91}, 023627 (2015).
% % %
\bibitem{degrandi}
C. De Grandi, V. Gritsev, and A. Polkovnikov, Phys. Rev. B {\bf 81}, 224301 (2010).
%
\bibitem{tezuka2014}
M. Tezuka, A. Garc\'{\i}a-Garc\'{\i}a, and M. A. Cazalilla, 
Phys. Rev. A {\bf 90},  053618 (2014).
%
\bibitem{cazalilladis}
M. A. Cazalilla, F. Sols, and F. Guinea, Phys.~Rev.~Lett. {\bf  97}, 076401 (2006). 
%
\bibitem{lobos}
A.~M. Lobos, M. Tezuka, and A. M. Garc\'{\i}a-Garc\'{\i}a,
Phys.~Rev. B {\bf 88}, 134506 (2013).
%
\bibitem{iontraps}
J. W. Britton, B. C. Sawyer, A. C. Keith, C.-C. J. Wang, J. K.
Freericks, H. Uys, M. J. Biercuk, and J. J. Bollinger, Nature (London) {\bf 484}, 489 (2012); R. Islam, C. Senko, W. C. Campbell, S. Korenblit, J. Smith, A. Lee, E. E. Edwards, C.-C. J. Wang, J. K. Freericks, and C. Monroe, Science {\bf 340}, 583 (2013).
%
\bibitem{rigol_eth1}
M. Rigol, Vanja Dunjko, and 
M. Olshanii, Nature (London) {\bf 452}, 854 (2008).
%
\bibitem{rigol_eth2}
M. Rigol,  arXiv:1511.04447 (2015).
%
\bibitem{diehl}
M. Buchhold, M. Heyl, and S. Diehl,
report arxiv:1510.03447 (2015).
%
\bibitem{yang2002}
Y. Yang, Y. Chung, D. Sprinzak, M. Heiblum, D. Mahalu, and H. Shtrikman, Nature (London) {\bf 422}, 415 (2002).
%
\bibitem{chalker1}
D.~L. Kovrizhin and J.~T. Chalker, Phys. Rev. B {\bf 84},
085105 (2011).
%
\bibitem{chalker2}
 D.~L. Kovrizhin and J.~T. Chalker, 
Phys.~Rev.~Lett. {\bf 109}, 106405 (2012).
%  
\bibitem{milletari}
M. Milletari and B. Rosenow, Phys.~Rev.~Lett.
{\bf 111}, 136807 (2013).
%
\bibitem{inue}
H. Inoue, A. Grivnin, N. Ofek, I. Neder, M. Heiblum, V. Umansky, and D. Mahalu, Phys. Rev. Lett. {\bf 112}, 166801 (2014).
%
%
\bibitem{eneko}
E. Malatsetxebarria, Z. Cai, U. Schollw\"ock, and M. A. Cazalilla, 
Phys. Rev. A {\bf 88}, 063630 (2013); Z. Cai, U. Sch\"ollwock, L. 
Pollet Phys. Rev. Lett. {\bf 113}, 260403 (2014).
%

\end{thebibliography}
\end{document}